\documentclass[prd,preprint,superscriptaddress,preprintnumbers,eqsecnum,showpacs,nofootinbib,nobibnotes]{revtex4}
\usepackage{amsfonts,amsmath,amssymb,bm,natbib}
\usepackage{graphicx} 
\usepackage{pstricks}

\newcommand{\be}{\begin{equation}}
\newcommand{\bea}{\begin{eqnarray}}
\newcommand{\ee}{\end{equation}}
\newcommand{\eea}{\end{eqnarray}}
\def\s#1{{\scriptscriptstyle #1}}

\def\1eq#1{Eq.~(\ref{#1})}

\def\2eqs#1#2{Eqs.~(\ref{#1}) and~(\ref{#2})}
\def\3eqs#1#2#3{Eqs.~(\ref{#1}),~(\ref{#2}) and~(\ref{#3})}
\def\noeq#1{(\ref{#1})}

\def\fig#1{Fig.~\ref{#1}}

\def\gb{\bm{\Gamma}}
\def\g{\widetilde\gb}

\def\diff#1{{\rm d}#1\,}

\def\ie{{\it i.e.}, }
\def\eg{{\it e.g.}, }

\def\gtree{\Gamma^{(0)}}
\def\gtreeb{\widetilde{\Gamma}^{(0)}}
\def\gfullb{\widetilde{\Gamma}}

\def\qqq{\Gamma}
\def\bqqm{\gfullb_{m}}
\def\qqqm{\qqq_{m}}
\def\Deltam{\Delta_{m}}
\def\Vbqq{\widetilde{V}}
\def\Vqqq{V}

\def\bcj{J}

\def\NV{\gfullb'}

\def\NP{\widetilde{V}}
\def\Jm{J_m}

\def\s#1{{\scriptscriptstyle #1}}
\def\E{\s{\mathrm{E}}}

\def\n#1{({\it #1}\,)}

\def\gA{g^2 C_A}
\def\gAsq{g^4 C^2_A}
\def\Y{Y}

\begin{document}

\title{The all-order equation of the effective gluon mass}


\author{D. Binosi}
\affiliation{European Centre for Theoretical Studies in Nuclear
Physics and Related Areas (ECT*) and Fondazione Bruno Kessler, \\Villa Tambosi, Strada delle
Tabarelle 286, 
I-38123 Villazzano (TN)  Italy}

\author{D. Iba\~nez}

\author{J. Papavassiliou}
\affiliation{\mbox{Department of Theoretical Physics and IFIC,  
University of Valencia}
E-46100, Valencia, Spain}

\begin{abstract}

We  present  the  general  derivation  of  the  full  non-perturbative
equation that governs 
the momentum evolution of the  dynamically generated gluon
mass, in  the Landau  gauge. 
The entire construction hinges crucially on the inclusion of longitudinally coupled
vertices containing massless poles 
of  non-perturbative origin, which  preserve the form  of the
fundamental  Slavnov-Taylor identities  of  the theory.  
The mass equation is obtained from a  previously unexplored version of the 
Schwinger-Dyson equation for the gluon propagator, 
particular  
to the PT-BFM formalism, which involves a reduced  
number of ``two-loop dressed'' diagrams, thus simplifying 
the calculational task considerably. 
The two-loop contributions  
turn out to be of paramount importance, modifying the qualitative features of the 
full mass equation, and enabling the emergence of physically meaningful solutions. 
Specifically, 
the resulting homogeneous 
integral equation is solved numerically, subject to certain approximations, for 
the entire range of physical momenta, yielding positive-definite 
and monotonically decreasing gluon masses.

\end{abstract}


\maketitle

\section{Introduction}
The dynamical generation of a momentum-dependent gluon mass~\cite{Cornwall:1981zr}
has received particular attention lately, 
especially in light of the important results obtained 
from a large number of lattice simulations. 
Specifically, 
the observed 
infrared (IR) finiteness of the  gluon
propagator and  the  ghost   dressing  function (in the Landau gauge), 
both in  $SU(2)$~\cite{Cucchieri:2007md,Cucchieri:2010xr} and  in $SU(3)$~\cite{Bogolubsky:2007ud,Bogolubsky:2009dc,Oliveira:2009eh},  
may be explained  in terms of such a nonperturbative mass~\cite{Aguilar:2008xm,RodriguezQuintero:2010ss,Pennington:2011xs}, 
which tames the IR divergences of the Green's functions of the theory (for alternative explanations, see,\eg \cite{Dudal:2008sp}).

The Schwinger-Dyson equations (SDEs) 
constitute the most natural framework for studying
such a nonperturbative phenomenon in the continuum~\cite{Binosi:2007pi,Binosi:2008qk,Alkofer:2000wg,Fischer:2006ub,Braun:2007bx}.
In particular,  
the emergence of 
IR finite solutions out of the SDE governing the gluon propagator
has been studied in detail in a series of works~(see \eg~\cite{Aguilar:2008xm,Fischer:2008uz,RodriguezQuintero:2010wy}). 
Particularly interesting in this context is the question of  
how to isolate the integral equation that determines the momentum evolution of 
the gluon mass~\cite{Aguilar:2011ux}. 
The main purpose of this work is to present the general derivation of the complete gluon mass equation,  
employing the {\it full} SDE of the gluon propagator.

In order to address this difficult question, we 
consider the set of modified SDE equations obtained within 
the general framework arising from the  synthesis of 
the pinch technique (PT)~\cite{Cornwall:1981zr,Cornwall:1989gv,Binosi:2002ft,Binosi:2003rr,Binosi:2009qm}  
with the background field method (BFM)~\cite{Abbott:1980hw},  
known in the literature as the PT-BFM scheme~\cite{Aguilar:2006gr,Binosi:2007pi,Binosi:2008qk}.
This becomes possible by virtue of set of powerful 
relations, known as the background-quantum identities~\cite{Grassi:1999tp,Binosi:2002ez}. These all-order relations 
allow one to express the conventional gluon propagator connecting two quantum ($Q$) gluons 
in terms of the two additional gluon propagators appearing in the BFM: 
the propagator connecting two background ($B$) gluons, and the propagator connecting 
a quantum with a background gluon ($BB$ and $QB$ propagators, respectively).
This, in turn, permits one to use the SDE for 
the $BB$ or $QB$ propagators, which are expressed in terms of the Feynman rules characteristic of the BFM,
involving fully dressed vertices that satisfy  Abelian-like Ward identities (WIs), 
instead of the typical Slavnov-Taylor identities (STIs). 
The main consequence of this formulation is that the resulting SDE may be suitably truncated, 
without compromising the transversality of the answer.
As we will explain in detail in the main body of the article, one gains a considerable 
advantage by considering the version of the identity connecting the $QQ$ with 
the $QB$ propagators, instead of the one connecting the $QQ$ with the $BB$ propagators, used in the literature so far. Specifically, the resulting SDE (to be denoted as the ``$QB$ version''
displays the powerful block-wise transversality 
property known from the  $BB$ case, but has two additional 
important features: it contains fewer graphs, and the limiting procedure 
necessary for projecting the result to the Landau gauge is significantly less involved.

Even within this improved framework, one still faces the fundamental question of how to disentangle  
from the SDE of the entire gluon propagator the part that determines the evolution of the mass 
from the part that controls the evolution of the ``kinetic'' term. 
In this work we present a new unambiguous way for implementing this separation, which exploits to the fullest  
the characteristic structure of a certain type of vertices that are inextricably connected with the 
process of gluon mass generation.    

Specifically, a crucial condition for  obtaining out of the SDEs 
an IR-finite gluon propagator,  without interfering with the 
gauge invariance (or the BRST symmetry) 
of the theory, is the existence of a set of special vertices, to be generically denoted by $V$, 
that are purely  longitudinal and contain massless poles, and 
must be added to the usual (fully-dressed) vertices of the theory.
The role of these vertices is two-fold. On the 
one hand, thanks to the massless poles they contain, they make possible the emergence of a 
IR finite solution out of the SDE governing the gluon propagator; thus, one invokes   
essentially a non-Abelian realization of the well-known Schwinger mechanism~\cite{Schwinger:1962tn,Schwinger:1962tp}.  
On the other hand, these same poles act like composite Nambu-Goldstone  excitations,  
preserving the form of the STIs of the theory 
in the presence of a gluon mass.
Recent studies indicate that 
the strong Yang-Mills dynamics can indeed generate  
longitudinally-coupled 
composite (bound-state) 
massless poles, which subsequently give rise to the required vertices  $V$~\cite{Aguilar:2011xe}. 

It turns out that the very nature of these vertices furnishes a solid guiding principle 
for implementing the 
aforementioned separation between mass and kinetic terms. In particular, 
their longitudinal structure, coupled to the fact that we work in the Landau gauge, 
completely determines the $q_{\mu}q_{\nu}$ component of the mass equation; this is tantamount to 
knowing the full mass equation, given that the answer is transverse (so, the $g_{\mu\nu}$ part is 
automatically fixed from its $q_{\mu}q_{\nu}$ counterpart). 
If, instead, one had tried to determine the $g_{\mu\nu}$ part first 
(or had taken the trace of the full equation, as was done in~\cite{Aguilar:2011ux}), 
one would have been confronted with a 
much subtler exercise. Specifically, the cancellation of the quadratic divergences, 
implemented by virtue of a characteristic identity, leads to a nontrivial mixing between 
the   $g_{\mu\nu}$ component of the mass and kinetic terms. Even though this separation can be eventually 
carried out, it is significantly more involved and delicate 
than that of the $q_{\mu}q_{\nu}$ components. 

As already mentioned, in the present work we include all fully dressed graphs 
(one and two loops) comprising the corresponding full SDE of the $QB$ propagator. 
Going beyond the ``one-loop'' dressed analysis is highly nontrivial, 
because it requires the introduction of a new $V$-type vertex, never   
considered before. Specifically, in addition to the $V$s related to the three-gluon vertices, 
known from the one-loop case,  
the $V$ vertex associated with the fully-dressed four gluon vertex $BQ^3$  
must be included in the corresponding ``two-loop dressed'' diagram. 
As happens in the one-loop case with the three-gluon  $V$, 
this new four-gluon $V$ vertex must satisfy a very concrete QED-like WI, 
in order to ensure the transversality of the ``two-loop dressed'' part of the calculation. 
Interestingly enough, and again as a consequence of their longitudinal nature and the 
Landau gauge, the WIs (and in some case the STI) 
satisfied by all $V$-type vertices involved in this problem 
is all that one needs for calculating their effects exactly. 
This fact clearly constitutes an important simplification and bypasses the need to actually 
construct explicitly the corresponding vertices.

As a consequence of the novel aspects introduced in our approach, 
the one-loop calculation presented in the first part 
of our derivation (Section~\ref{oneloop}) proceeds in a far more concise way 
compared to the corresponding derivation followed in~\cite{Aguilar:2011ux},  
rectifying, in fact, the form of the resulting mass equation. 
The two-loop contribution is considerably more cumbersome to obtain, 
and is expressed in terms of a kernel that,  
in addition to full gluon propagators, 
involves also the conventional, fully dressed, three gluon vertex ($Q^3$).
The two-loop part of the mass equation is subsequently simplified 
by choosing tree-level values for a judicious combination of its ingredients, 
a fact that allows us to carry out explicitly one of the two integrations over virtual momenta.
In order to gain insight on the numerical subtleties associated with this equation, 
we first consider its limit at vanishing physical momenta, 
thus converting it into a nonlinear constraint.
Already at this level, the contribution from the two-loop part 
appears to be of paramount importance, having 
far-reaching consequences for the behavior of the resulting solutions.
The detailed numerical 
solution of the full equation (for arbitrary values of the physical momentum) 
confirms this impression, revealing the existence of 
positive-definite and monotonically decreasing solutions.

The article is organized as follows.
In Section~\ref{QBsec} we introduce the basic notation, together with some of the 
most important PT-BFM relations, and 
explain the derivation and general structure of the $QB$ version of the SDE for the gluon propagator. 
In Section~\ref{remind}  we present a brief reminder of the basic elements appearing in the 
gauge-invariant generation of a gluon mass, with particular emphasis on the 
role and properties of the $V$ vertices.
In Section~\ref{genmet} we outline in detail the methodology that allows one to extract from the 
corresponding SDE  
two separate equations, one for the mass and one for the kinetic term, 
and explain the advantage of selecting out the $q_{\mu}q_{\nu}$ component.   
Section~\ref{oneloop} is dedicated to the concise derivation of the one-loop version of the 
mass equation. In addition, we explain in passing the reason for the discrepancy with the 
result given in~\cite{Aguilar:2011ux}; in addition, 
a brief discussion on the ghost sector 
(only ``one-loop'' in this new SDE version) is included, 
explaining why the ghost graphs do not affect the mass equation. In Section~\ref{twoloop}  
we present the full two-loop calculation, organizing the 
corresponding technical aspects into various self-contained subsections. 
In Section~\ref{fullmasseq} we present the final form of the 
integral equation that governs the dynamical mass, and   
discuss some of its general properties. In addition, we calculate an approximate expression for the 
new contribution to the kernel of this integral equation, to be used in the ensuing numerical analysis.    
In Section~\ref{numan} we solve numerically the integral equation,  
determine a family of 
positive-definite and monotonically decreasing solutions, and study their dependence on the 
value of the strong coupling constant.
Finally, our conclusions are presented in Section~\ref{concl}.

\section{\label{QBsec}A simpler version of the fundamental SDE}

In a general renormalizable $R_\xi$ gauge, defined through 
a linear gauge-fixing function of the Lorentz type (${\cal F}^a=\partial^\mu A^a_\mu$), 
the all-order gluon propagator $\Delta^{ab}_{\mu\nu}(q)=\delta^{ab}\Delta_{\mu\nu}(q)$ 
and its inverse read
\be
i\Delta_{\mu\nu}(q)=- i\left[P_{\mu\nu}(q)\Delta(q^2)+\xi\frac{q_\mu q_\nu}{q^4}\right];\qquad
\Delta^{-1}_{\mu\nu}(q)=i\left[P_{\mu\nu}(q)\Delta^{-1}(q^2)+ (1/\xi)q_\mu q_\nu\right],
\label{prop}
\ee 
where  $\xi$ denotes the  gauge-fixing parameter ($\xi=0$ corresponds to the Landau  gauge), and 
\be  P_{\mu\nu}(q)=g_{\mu\nu}- q_\mu q_\nu/q^2 \,,
\ee 
is the  dimensionless transverse  projector.  The scalar  form factor
$\Delta(q^2)$  appearing  above  is  related to  the  all-order  gluon
self-energy $\Pi_{\mu\nu}(q)$;  this latter  quantity, as a consequence of the BRST symmetry, 
is transverse to all orders in perturbation  theory, as well as non perturbatively, at
the   level   of   the   corresponding   SDE.   One   has   then   
\be
q^\mu\Pi_{\mu\nu}(q)=0;\qquad \Pi_{\mu\nu}(q)=\Pi(q^2)P_{\mu\nu}(q).
\label{trangen}
\ee

The nonperturbative dynamics of the gluon propagator in the continuum is governed by the corresponding SDE. 
As has been explained in detail in a series of articles, 
the formulation of this dynamical equation in the context of the  PT-BFM formalism furnishes certain distinct  
advantages over the conventional case; most importantly,
it permits the systematic truncation of the SDE 
series, without compromising the BRST symmetry, as captured by (\ref{trangen}).

\begin{figure}[!t]
\includegraphics[scale=.6]{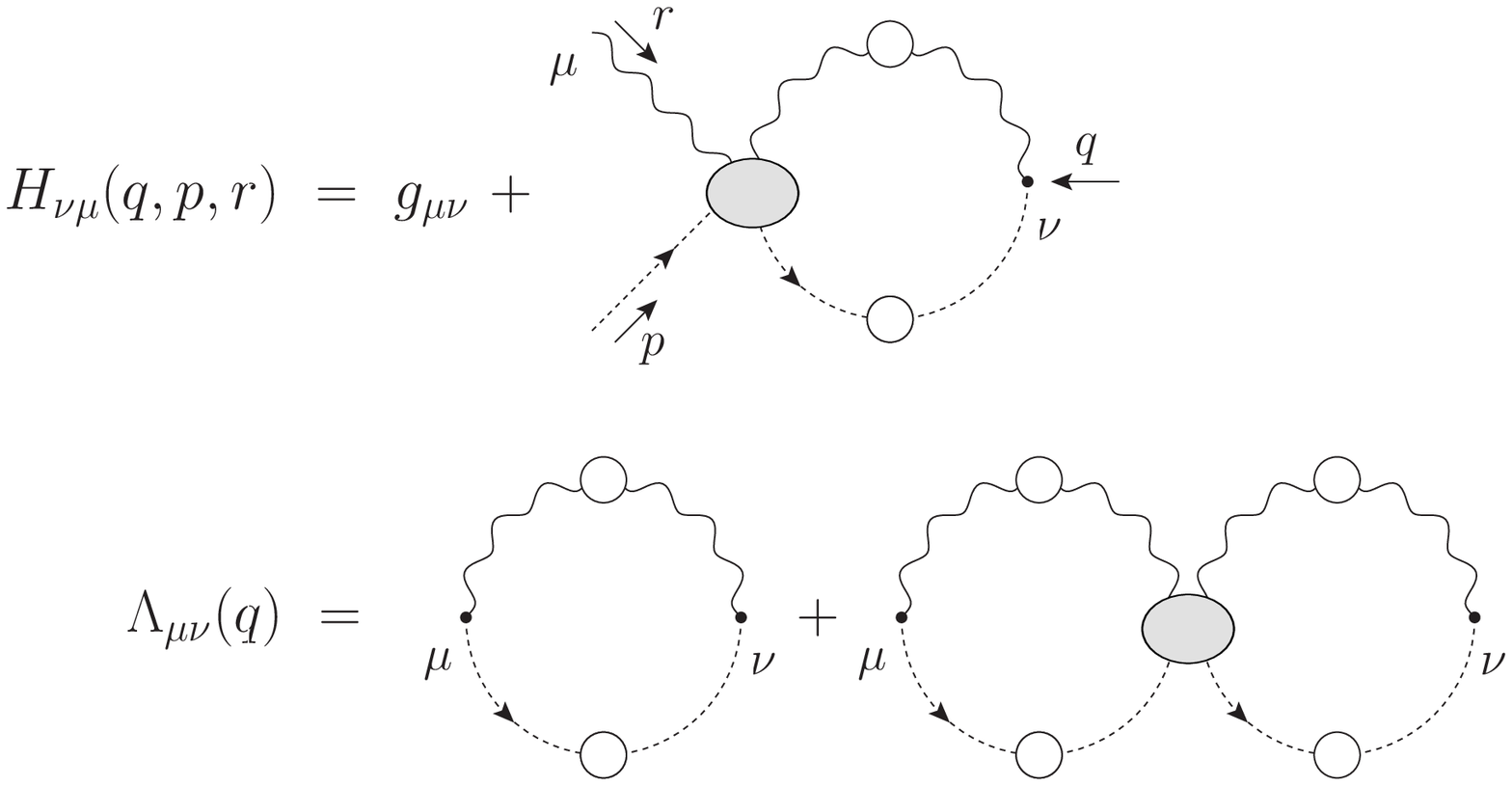}
\caption{\label{H-Lambda-new}Definitions and conventions of the auxiliary functions $\Lambda$ and $H$. 
The color and gauge-coupling dependence 
for the  field combination $c^a(p)A_\mu^b(r)A_\nu^{*c}(q)$ shown in the $H$ function is $gf^{acb}$.  White blobs denote connected Green's functions, while gray blobs denote  
one-particle irreducible (with respect to vertical cuts) Schwinger-Dyson kernels.}
\end{figure}

Within the BFM formalism three types of gluon propagator make their appearance, in a natural way: 
({\it i}) the conventional gluon propagator (two quantum gluons entering, $QQ$), 
denoted (as above) by $\Delta(q^2)$; ({\it ii})
the background  gluon propagator (two background gluons entering, $BB$), 
denoted by $\widehat\Delta(q^2)$; and ({\it iii})
the mixed  background-quantum gluon propagator (one background and one quantum gluons entering, $BQ$), 
denoted by  $\widetilde\Delta(q^2)$. These three propagators are related  among each other
by a set of powerful relations, 
known as Background-Quantum identities~\cite{Grassi:1999tp, Binosi:2002ez},  
obtained within the Batalin-Vilkovisky formalism~\cite{Batalin:1977pb,Batalin:1981jr}. 
Specifically,  
\bea
\Delta(q^2) &=& [1 + G(q^2)]^2 \widehat{\Delta}(q^2),
\nonumber\\
\Delta(q^2) &=& [1 + G(q^2)] \widetilde{\Delta}(q^2), 
\nonumber\\
\widetilde{\Delta}(q^2) &=& [1 + G(q^2)] \widehat{\Delta}(q^2).
\label{BQIs}
\eea
The function $G(q^2)$, which is instrumental for enforcing these crucial relations, is defined as the $g_{\mu\nu}$ form factor of a special two-point function, given by (see Fig.~\ref{H-Lambda-new})
\bea
\Lambda_{\mu\nu}(q)&=&-i\gA\int_k\!\Delta_\mu^\sigma(k)D(q-k)H_{\nu\sigma}(-q,q-k,k)\nonumber\\
&=&g_{\mu\nu}G(q^2)+\frac{q_\mu q_\nu}{q^2}L(q^2),
\label{Lambda}
\eea
where $C_A$ represents the Casimir eigenvalue of the adjoint representation ($N$ for $SU(N)$), 
$d=4-\epsilon$ is the space-time dimension, and we have introduced the integral measure
\be
\int_{k}\equiv\frac{\mu^{\epsilon}}{(2\pi)^{d}}\!\int\!\mathrm{d}^d k,
\label{dqd}
\ee
with $\mu$ the 't Hooft mass. In addition, 
$D^{ab}(q^2)=\delta^{ab}D(q^2)$ is the ghost propagator, and $H_{\nu\sigma}$ is 
the gluon-ghost kernel shown.
The dressed loop expansion of $\Lambda$ and $H$ is shown in~\fig{H-Lambda-new}; notice finally that, in the Landau gauge, an important all-order relation exists, which 
links the form factors $G(q^2)$ and  $L(q^2)$ to the 
ghost dressing function $F(q^2)=q^2D(q^2)$, 
namely~\cite{Grassi:2004yq,Aguilar:2009nf,Aguilar:2009pp,Aguilar:2010gm}  
\be
F^{-1}(q^2) = 1 + G(q^2) + L(q^2).
\label{funrel}
\ee

The basic observation put forth in~\cite{Aguilar:2006gr,Binosi:2007pi,Binosi:2008qk}
is that one may use the SDE for $\widehat{\Delta}(q^2)$, written in terms of the 
BFM Feynman rules, take advantage of its improved truncation properties,  and then 
convert it to an equivalent equation for  $\Delta(q^2)$ (the propagator simulated on the lattice) 
by means of the first relation in \1eq{BQIs}. The resulting SDE contains a richer diagrammatic 
structure than the conventional one, due to the appearance of a new subset of graphs, 
related to the modified ghost sector characteristic of the BFM; specifically, one encounters 
a new type of ghost vertex, involving two gluons and two ghosts (for the complete set of all relevant Feynman rules, see, for example,~\cite{Binosi:2008qk}). 

\begin{figure}[!t]
\includegraphics[scale=1.1]{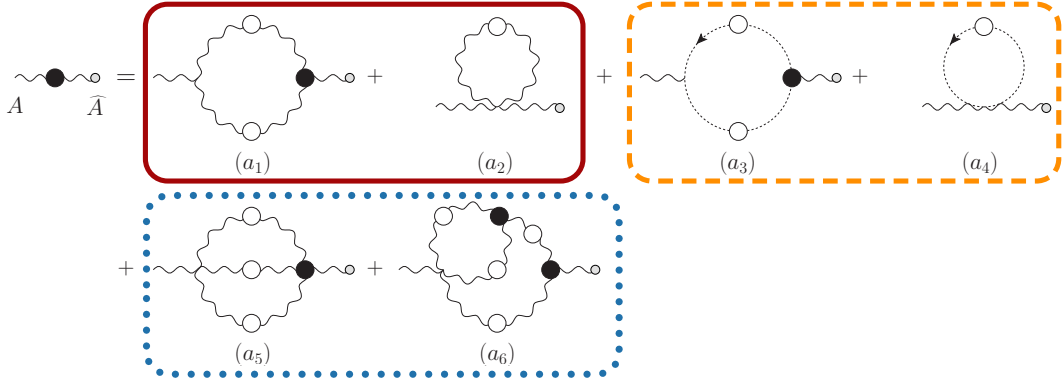}
\caption{\label{glSDE}(color online). The SDE obeyed by the $QB$ gluon propagator. 
Each of the three different boxes (continuous, dashed and dotted line) encloses a set of diagrams forming a transverse subgroup. 
Black blobs represents fully dressed 1-PI vertices; 
the small gray circles appearing on the external legs (entering from the right, only!) are used to indicate background gluons.}
\end{figure}

In this article we present an alternative version of the gluon SDE, 
which displays all known desirable features, and, at the same time, has a reduced diagrammatic complexity.
Specifically, instead of using $\widehat{\Delta}(q^2)$ as the starting point, we consider the 
SDE satisfied by $\widetilde{\Delta}(q^2)$, shown in Fig.\ref{glSDE}, and the second (instead of the first) relation 
in \1eq{BQIs}. 
Then, the corresponding version of the SDE for the conventional gluon propagator (in the Landau gauge) reads  
\be
\Delta^{-1}(q^2){ P}_{\mu\nu}(q) = 
\frac{q^2 {P}_{\mu\nu}(q) + i\,\sum_{i=1}^{6}(a_i)_{\mu\nu}}{1+G(q^2)},
\label{sde}
\ee
where the diagrams $(a_i)$ are shown in Fig.~\ref{glSDE}. 

The crucial points to recognize are \n{i} one has a reduced set of Feynman diagrams (six instead of ten)
compared to those appearing in the previous formulation in terms of $\widehat{\Delta}(q^2)$
(six instead of ten, respectively), and \n{ii}  
the most important feature of the PT-BFM formalism for 
our purposes, namely the block-wise transversality imposed at the level of the SDE for the gluon self-energy, is still 
present. In particular, 
the transversality of the gluon self-energy is realized  according to the pattern highlighted by the boxes of Fig.~\ref{glSDE},
namely, 
\be
q^{\mu} [(a_1) + (a_2)]_{\mu\nu} = 0;\qquad
q^{\mu} [(a_3) + (a_4)]_{\mu\nu} = 0; \qquad
q^{\mu} [(a_5) + (a_6)]_{\mu\nu} = 0.
\label{boxtr2}
\ee 

\section{\label{remind}Gluon mass generation: a brief reminder}

As has been explained in detail in the recent literature, the self-consistent gauge-invariant generation of a gluon mass 
proceeds through the implementation of the well-known Schwinger mechanism  in the context of a Yang-Mills theory.
This mechanism 
requires the existence of a very special type of nonperturbative vertices, to be generically denoted by $V$
(with appropriate Lorentz and color indices], which, when added 
to the conventional fully dressed vertices have a triple effect: 
\n{i} they make possible that the SDE of the gluon propagator yields $\Delta^{-1}(0)\neq 0$;
\n{ii} they guarantee that the WIs and STIs of the theory 
remain intact, \ie they maintain exactly the same form before and after mass generation; and 
\n{iii} they decouple from {\it on-shell} amplitudes. 
These three properties become possible because the special vertices \n{a}
contain massless poles 
and \n{b} are  
completely {\it longitudinally} coupled, \ie they satisfy 
conditions such as (for the case of the three-gluon vertex) 
\be
P^{\alpha'\alpha}(q) P^{\mu'\mu}(r) P^{\nu'\nu}(p) \Vbqq_{\alpha'\mu'\nu'}(q,r,p)  = 0.
\label{totlon}
\ee   

The origin of the aforementioned poles is due to purely non-perturbative dynamics: for sufficiently strong binding, 
the mass of certain (colored) bound states 
may be reduced to zero~\cite{Jackiw:1973tr,Jackiw:1973ha,Cornwall:1973ts,Eichten:1974et,Poggio:1974qs}.  
In addition to  triggering the Schwinger mechanism, these bound-state poles 
act as composite, longitudinally coupled Nambu-Goldstone bosons, maintaining gauge invariance. Notice, however, that they differ from ordinary Nambu-Goldstone bosons 
as far as their origin is concerned, since they are not associated 
with the spontaneous breaking of any continuous symmetry~\cite{Cornwall:1981zr}.
This dynamical scenario has been shown to be realized, within a set of simplified assumptions, through a detailed study in~\cite{Aguilar:2011xe}.

For the ensuing analysis 
it is advantageous to introduce the {\it inverse} 
of the gluon dressing function, to be 
denoted by  $J(q^2)$. Specifically, in the absence of a gluon mass, we write 
\be
\Delta^{-1}({q^2})=q^2 J(q^2).
\label{defJ}
\ee
From the kinematic point of view 
we will describe the transition 
from a massless to a massive gluon propagator by carrying out the replacement  
(in Minkowski space)
\be
\Delta^{-1}(q^2) = q^2 J(q^2) \quad \longrightarrow\quad  \Deltam^{-1}(q^2)=q^2 \Jm(q^2)-m^2(q^2),
\label{massive}
\ee
where $m^2(q^2)$ is the (momentum-dependent) dynamically generated mass, and the subscript ``$m$'' in 
$\Jm$ indicates that, effectively, one has now a mass inside the corresponding expressions 
(\ie in the SDE graphs).

Gauge invariance requires that the replacement described schematically in \1eq{massive} 
be accompanied by a simultaneous replacement of all relevant vertices by 
\be
\gfullb \quad \longrightarrow\quad   \NV = \bqqm + \Vbqq ,  
\label{nv}
\ee
where $\Vbqq$ must be such that the new vertex $\NV$ 
satisfies the same WIs (or STI) as $\gfullb$, but now replacing the 
gluon propagators appearing on their rhs by massive ones.

To see how this works with an explicit example, introducing at the same time  some 
necessary ingredients for the analysis that follows,   
consider the fully dressed vertex $BQ^2$, 
connecting a background gluon with two quantum gluons, 
to be denoted by $\widetilde{\Gamma}_{\alpha\mu\nu}$
With the Schwinger mechanism ``turned off'', this vertex satisfies the WI
\be
q^\alpha\widetilde{\Gamma}_{\alpha\mu\nu}(q,r,p)=p^2\bcj(p^2)P_{\mu\nu}(p)-r^2\bcj(r^2)P_{\mu\nu}(r),
\label{STI}
\ee
when contracted with respect to the momentum of the background gluon.
The general replacement described in (\ref{nv}) amounts to introducing the vertex  
\be
\NV_{\alpha\mu\nu}(q,r,p) = \left[\widetilde{\Gamma}_{m}(q,r,p) + \NP(q,r,p)\right]_{\alpha\mu\nu};
\label{NV}
\ee
then,  gauge invariance requires that 
\be
q^\alpha \NP_{\alpha\mu\nu}(q,r,p)= m^2(r^2)P_{\mu\nu}(r) - m^2(p^2)P_{\mu\nu}(p),
\label{winp}
\ee
so that, after turning the Schwinger mechanism on,  the corresponding WI satisfied by $\NV$ would read  
\bea
q^{\alpha}\NV_{\alpha\mu\nu}(q,r,p) &=& 
q^{\alpha}\left[\widetilde{\Gamma}_{m}(q,r,p) + \NP(q,r,p)\right]_{\alpha\mu\nu}
\nonumber\\
&=& [p^2 \Jm (p^2) -m^2(p^2)]P_{\mu\nu}(p) - [r^2 \Jm (r^2) -m^2(r^2)]P_{\mu\nu}(r)
\nonumber\\
&=& \Deltam^{-1}({p^2})P_{\mu\nu}(p) - \Deltam^{-1}({r^2})P_{\mu\nu}(r) \,,
\label{winpfull}
\eea
which is indeed the identity in Eq.~(\ref{STI}), with the aforementioned replacement 
\mbox{$\Delta^{-1} \to \Delta_m^{-1}$} enforced. 
The remaining STIs, triggered when contracting $\NV_{\alpha\mu\nu}(q,r,p)$ 
with respect to the other two legs  
are realized in exactly the same fashion.

Finally, note that ``internal'' vertices, namely vertices involving only quantum gluons,  
must be also supplemented by the corresponding $V$, such that their STIs are also unchanged 
in the presence of masses. To be sure, these  
vertices do not contain $1/q^2$-type of poles, but rather poles in the virtual momenta;
therefore, they cannot contribute directly to the mass-generating mechanism. However, 
they must be included anyway, for gauge invariance to remain intact.  

Before concluding this section, it is important to clarify some additional points  
related to the concepts presented here.

To begin with,    
a sharp distinction between 
the notions of ``gauge invariance'' and ``gauge independence'' must be established. Gauge invariance 
is used throughout this work 
for indicating that a Green's function satisfies the WI (or STI) 
imposed by the gauge (or BRST) symmetry of the theory. On the other hand, the gauge (in)dependence 
of a Green's function is related with the (independence of) dependence on the gauge-fixing 
parameter (\eg $\xi$) used to quantize the theory. Evidently, an off-shell Green's function may be 
gauge invariant but gauge dependent: for example, the QED photon-electron vertex, $\Gamma_{\mu}(p,p+q)$,  
depends explicitly on $\xi$ but satisfies (for every value of  $\xi$) the 
classic WI   \mbox{$q^{\mu} \Gamma_{\mu}(p,p+q) = S^{-1}(p+q) - S^{-1}(p)$}. 
A text-book example of a Green's function that is both gauge invariant 
and gauge independent is the photon self-energy (vacuum polarization), which is both 
transverse and $\xi$ independent.

Evidently, the procedure followed for obtaining the 
the gluon mass reported in this article (Section \ref{numan}) is gauge invariant, 
because,  as explained above, the presence of the vertices $V$ 
maintains the STIs intact, and the transversality of the $\Pi_{\mu\nu}(q)$ is guaranteed.
However, it is clear that 
a particular gauge choice has been implemented from the beginning, namely that of the Landau gauge.
Thus, the gluon mass so derived is particular to that gauge, and we are not aware 
of any arguments supporting the notion  
that the {\it same} mass would emerge if the SDE analysis  
were to be repeated, for example, in the Feynman gauge. 
Therefore, the Landau gauge mass found here 
should not be considered as an ``observable'', in the strict sense of the term.
In fact, a qualitatively different picture may appear in the context 
of  different gauge-fixing schemes,  
such as the Coulomb gauge or the maximal Abelian gauge~\cite{Kronfeld:1987ri};  for instance, in the former 
gauge, only ``scaling'' solutions (for which $\Delta(0)=0$) have been reported~\cite{Burgio:2009xp,Watson:2011kv}.  

\begin{figure}[!t]
\includegraphics[scale=.6]{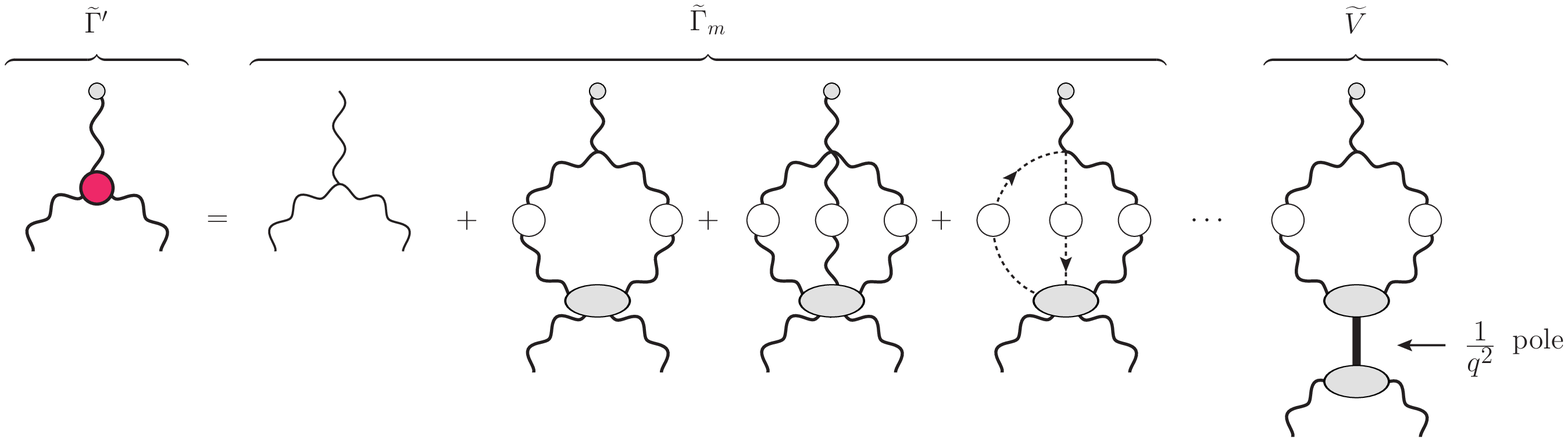}
\vspace{-1.8cm}
\caption{\label{Gammaprime}(color online). The $\NV$ three-gluon vertex. 
Thick  gluon lines indicate massive gluons.}
\end{figure}

The mass-generation mechanism employed here (and previous works) 
hinges crucially on the presence of the special vertices $V$, with the main properties described 
above. 
As has been explained in ~\cite{Jackiw:1973tr,Jackiw:1973ha,Cornwall:1973ts,Eichten:1974et,Poggio:1974qs,Aguilar:2011xe}, 
their emergence proceeds through 
the dynamical formation and subsequent exchange of a massless composite excitation; 
the massless pole corresponds to the propagator of this excitation.  
This new amplitude modifies the 
usual skeleton expansion accordingly, as shown in~\fig{Gammaprime}, 
given that conventional vertices do not possess such poles.  
Thus, in that sense, one may interpret the origin of the $V$ vertices 
as being due to dynamical non-localities in the theory.  

\pagebreak

It is, therefore, important to recognize the following points.  

\begin{itemize}

\item[({\it a})] These non-localities do not enter at the level of the fundamental QCD 
Lagrangian, which remains unaltered. In fact, by composite excitation 
we mean a pole in an off-shell Green's function representing a field that 
does not exist in the classical action but that occurs in the solution of the SDE for that Green's function, as a sort of bound state.

\item[({\it b})] If one were to describe the $V$ vertices in terms of an effective Lagrangian, one would have 
indeed to introduce non-localities, as was done in~\cite{Cornwall:1981zr,Cornwall:1985bg}. 
The resulting theory corresponds to a non-linear sigma model that 
is non-renormalizable, and can only serve as an approximate low-energy description. 
We emphasize that we do not use any such model here nor anywhere in the cited works.

\item[({\it c})] Related to the previous point is the effect that the 
appearance of such vertices might have on the renormalizability of the 
Yang-Mills theory. It is clear that no divergences may be introduced that 
will be in any way proportional to the $V$ vertices, because their renormalization 
would then require the introduction of non-local terms, and one would effectively end up in 
the situation described at point ({\it b}). The preliminary study~\cite{Aguilar:2009ke,Binosi:2011wi,Aguilar:2011ux} 
of this issue suggests that the $V$ vertices 
give rise to the gluon mass, with no other residual effects. In fact, 
eventually, the field-theoretic quantities comprising  
these vertices are such that the renormalization proceeds 
identically as that of the conventional part, \ie through multiplication by the 
same renormalization constants employed in the massless case. 

\item[({\it d})] Finally, 
the dynamical equation describing the gluon mass [see \1eq{fullmass}] 
ought to be made finite only through appropriate multiplicative renormalization 
of its internal substructures, exactly as happens with 
the dynamical equation (gap equation) describing the 
momentum evolution of the constituent quark mass~\cite{Roberts:1994dr}. 
The validity of this important property will be assumed in 
Section~\ref{fullmasseq}, when dealing with \1eq{fullmass}.
Note also that, as has been explained in the early works on the subject~\cite{Cornwall:1981zr},
the renormalizability is intimately related to the 
vanishing of the gluon mass for large momenta (as seen in our solutions, see Fig.\ref{full-sols}). 
Without such large momentum falloff, the SDE would have solutions 
with extra infinities not corresponding to the perturbative renormalization principles.
\end{itemize}

\section{\label{genmet}Deriving the mass equation: General methodology}

As explained in the previous section, 
the special vertices $\widetilde V$ (and $V$) enforce  
the transversality of the rhs of (\ref{sde}) in the presence of gluon masses.
Specifically, writing the $\Deltam^{-1}(q^2)$ on the lhs of \1eq{sde}
in the form given in \1eq{massive}, 
one has that 
\be
[q^2 \Jm(q^2)-m^2(q^2)]{P}_{\mu\nu}(q) = 
\frac{q^2 {P}_{\mu\nu}(q) + 
i\, \sum_{i=1}^{6}(a_i^{\prime})_{\mu\nu}}{1+G(q^2)},
\label{sdem}
\ee
where the ``primes''  indicates that (in general) the various fully dressed vertices 
appearing inside the corresponding diagrams 
must be replaced by their primed counterparts, as in~\1eq{nv}. 
Thus, in graph $(a_1)$ the $BQ^3$ vertex will be substituted by $\NV$,  
while in graph $(a_6)$
both the $BQ^2$ and the $Q^3$ type of vertices must contain 
the corresponding $\NP$ and $V$ components, respectively. 
In addition, in diagram $(a_5)$  
the primed version of the vertex $BQ^3$ will make its appearance.
These modifications have an important effect: the blockwise transversality 
property of \1eq{boxtr2} holds also for the ``primed'' graphs, \ie when  
$(a_i)\to (a_i^{\prime})$. 

The lhs  of \1eq{sdem} 
involves two unknown quantities, $J_m(q^2)$ and $m^{2}(q^2)$, which will eventually 
satisfy two separate, but coupled, integral equations. 
of the generic type
\bea
J_m(q^2) &=& 1+ \int_{k} {\cal K}_1 (q^{2},m^2,\Delta_m),
\nonumber\\
m^{2}(q^2) &=&  \int_{k} {\cal K}_2 (q^{2},m^2,\Delta_m).
\label{separ}
\eea
such that $q^{2} {\cal K}_1 (q^{2},m^2,\Delta_m) \to 0$, as $q^{2}\to 0$, 
whereas ${\cal K}_2 (q^{2},m^2,\Delta_m)\neq 0$ in the same limit,
precisely because it includes the  $1/q^2$ terms contained inside the $\widetilde{V}$ terms.  

In the present work we will focus on the derivation of  
the closed form of the integral equation governing $m^{2}(q^2)$. 
To that end, we must identify all 
mass-related contributions contained 
in the Feynman graphs that comprise the rhs of \1eq{sdem}.
Now, with the transversality of both sides of \1eq{sdem} guaranteed, 
it turns out that it is far more economical to 
derive the mass equation by 
isolating the appropriate cofactors of 
$q_\mu q_\nu/q^2$ on both sides, instead of the $g_{\mu\nu}$, or instead of taking the trace. 
In particular, to obtain the rhs of the mass equation, 
one must \n{i} consider the   
graphs that contain a vertex ${\widetilde V}$,  
and \n{ii} isolate 
the $q_{\mu} q_{\nu} /q^{2}$ component of the 
contributions coming from the ${\widetilde V}$ vertices.

To explain how one reaches the above conclusion, 
let us first point out that, clearly, in the absence of the $\widetilde{V}$, which contain the massless poles, 
no gluon mass could be generated; so, the gluon mass is inextricably connected with the $\widetilde{V}$s.
In addition, as we will see in detail in what follows, 
the longitudinal nature of these latter vertices [{\it viz.} \1eq{totlon}], coupled to the 
fact that we work in the Landau gauge, force the corresponding contribution to the gluon self-energy to be proportional to 
$q_\nu$ [see \2eqs{1l-dr}{diagram3}]. The only exception to this rule is the 
$V$ that appears inside graph $(a_6)$, as part of the $Q^3$ vertex $\Gamma^{\prime}$;   
however, the 
corresponding contribution is shown to 
vanish identically in the Landau gauge (see Section~\ref{twoloop}). Thus, 
if we denote by $(a^{\s {\widetilde V}}_{i})_{\mu\nu}$ the 
${\widetilde V}$-related 
contributions of the corresponding diagrams, these latter terms are 
proportional to $q_\mu q_\nu/q^2$ only, namely
\be
(a^{\s {\widetilde V}}_{i})_{\mu\nu} = \frac{q_\mu q_\nu}{q^2} a^{\s {\widetilde V}}_{i}(q^2),
\ee
so that
\be
m^2(q^2) = \frac{ i \sum_{i}  a_{i}^{\s {\widetilde V}}(q^2)}{1+G(q^2)}\,,
\label{masseq}
\ee
where the sum includes only the graphs $i=1,5,6$.

Similarly, the equation for $J_m(q^2)$ 
may be obtained from the  $q_{\mu} q_{\nu} /q^{2}$ component 
of the parts of the graphs that do not contain $V$ components. These graphs 
are identical to the original set $(a_1)-(a_6)$, but now $\gfullb \longrightarrow\bqqm$, 
$\Delta \longrightarrow \Delta_m$, etc. To avoid notational clutter we will use the 
same letter as before, and the aforementioned changes are understood.
These contributions may be separated in  $g_{\mu\nu}$ and $q_{\mu} q_{\nu} /q^{2}$ components,  
\be
(a_i)_{\mu\nu} = g_{\mu\nu}\, {A_i}(q^2)  + \frac{q_{\mu} q_{\nu}}{q^{2}} {B_i}(q^2).
\ee
Note that graphs $(a_2)$ and $(a_4)$
are proportional to $g_{\mu\nu}$ only; so, in the notation introduced above, $B_2(q^2)=B_4(q^2)=0$.
Then, the corresponding equation for $J_m(q^2)$ reads 
\be
- q^2 \Jm(q^2) =  \frac{-q^2 + i \sum_{i} B_i(q^2)}{1+G(q^2)},
\label{sdeJ}
\ee
with $i=1,3,5,6$. 

We hasten to emphasize that the fact that we focus on the $q_\mu q_\nu/q^2$
terms, instead of the $g_{\mu\nu}$, in no way indicates a potential clash with  
the transversality of the gluon self-energy, which is manifestly preserved throughout.
In fact, it is precisely the validity of \1eq{trangen} that allows one to choose 
freely between the two tensorial structures. 
The point is that the transversality of \1eq{sdem}
should not be interpreted to mean that the algebraic origin of the terms  
proportional to $g_{\mu\nu}$ is the same as that of the terms proportional to $q_\mu q_\nu/q^2$.
In particular, the  $q_\mu q_\nu/q^2$  of  $\Jm(q^2)$ and $m^2(q^2)$ are easily separable, 
as the \2eqs{masseq}{sdeJ} indicate, whereas  their $g_{\mu\nu}$ parts are entangled, 
and their separation is significantly  more delicate. 

As a particular example of how the $g_{\mu\nu}$
part requires an elaborate treatment, while the $q_{\mu} q_{\nu} /q^{2}$ does not, 
let us consider the basic cancellation taking place at the one-loop dressed level,   
enforced by the so-called ``seagull identity'':  
the (quadratic) divergence of the ``seagull'' diagram ($a_2$) 
is annihilated 
exactly by a very particular contribution coming from graph ($a_1$), by virtue of the 
identity (valid in dimensional regularization)~\cite{Aguilar:2009ke}
\be
\int_k\! k^2 \frac{\mathrm{d}{\Delta_m (k^2)}}{{\mathrm d}k^2}+\frac{d}2\int_k\!\Delta_m(k^2)=0.
\label{seagull}
\ee 
The important fact to realize is that all terms involved in the cancellation enforced by \1eq{seagull}
are proportional to $g_{\mu\nu}$, and it is only after they are properly combined that 
their total contribution vanishes (as $q^{2} \to 0$), 
by virtue of \1eq{seagull}; instead, 
their $q_{\mu} q_{\nu} /q^{2}$ counterparts vanish {\it individually}, in the same limit.

In order to appreciate this last point, consider the expression   
\be
I_{\mu\nu}(q) \equiv \int_k\!k_\mu k_\nu f(k,q),
\ee
where $f(k,q)$ is an arbitrary function that remains finite in the limit $q^2 \to 0$.
Clearly, 
\be
I_{\mu\nu}(q) = g_{\mu\nu}{A}(q^2) + \frac{q_\mu q_\nu}{q^2} B(q^2)\,,
\ee
and the form factors $A(q^2)$ and $B(q^2)$ are given by 
\bea
A(q^2) &=& \frac1{d-1} \int_k\! \left[ k^2 - \frac{(k \cdot q)^2}{q^2}\right] f(k,q) ,
\nonumber\\
B(q^2)  &=& -\frac1{d-1} \int_k\! \left[ k^2 - d\frac{(k \cdot q)^2}{q^2}\right] f(k,q).
\label{justI}
\eea
Then, setting $(q\cdot k)^2 = q^2 k^2 \cos^2\theta$, and using that, for any function $f(k^2)$ 
\be
\int_k\!\cos^2\theta  f(k^2)=\frac1d\int_k\! f(k^2),
\label{costheta-rel}
\ee
we obtain from \1eq{justI} that, as $q^2 \to 0$, 
\be
A(0) = \frac1{d} \int_k\!k^2 \,f(k);\qquad B(0) =0.
\label{I0}
\ee
Evidently, the function $f$ may be such that the integral defining $A(0)$ diverges, while, for the same 
function, $B(0)$ vanishes. 

Now, in the context of the one-loop dressed calculation, the term $I_{\mu\nu}(q)$ 
originates from graph $(a_1)$, with the replacement $\gfullb \to\bqqm$, 
$\Delta \to \Delta_m$, as mentioned above. The WI satisfied by $\bqqm$ is that of \1eq{STI} 
with $J \to J_m$,
and similar but more complicated STIs hold when contracting with respect to the other momenta.
Note that what  appears in the WI and STIs is $J_m$ and {\it not} $\Delta_m$.
Then, as has been shown in \cite{Binosi:2011wi},  
the longitudinal part of $\bqqm$ may be expressed in terms of $J_m$ (and other Green's functions), and as a result,
the function $f(k)$  (at $q^2=0$) is given by 
\be
f(k^2) = -\Delta^2_m (k^2) \frac{{\mathrm d} \left(k^2 J_m (k^2)\right)}{{\mathrm d} k^2}.
\label{fk}
\ee 
Now, the point is that, in order to trigger \1eq{seagull}, $f(k^2)$ should be instead 
\be
f(k^2) = -\Delta^2_m (k^2) \frac{{\mathrm d}\Delta^{-1}(k^2)}{{\mathrm d} k^2} = \frac{{\mathrm d}\Delta_m (k^2)}{{\mathrm d} k^2}. 
\ee   
To accomplish this, one adds and subtracts $m^2(k^2)$ to the $f(k)$ of the $A(0)$ given in \1eq{I0},
thus obtaining 
\be
A(0) = \frac1{d}\left[ \int_k\!k^2 \frac{{\mathrm d}\Delta_m (k^2)}{{\mathrm d} k^2} -  \int_k\!k^2 \frac{{\mathrm d} m^2 (k^2)}{{\mathrm d} k^2}\right].
\label{I0mod}
\ee
At that point, the first term on the rhs of \1eq{I0mod} 
goes to the $g_{\mu\nu}$ part of the equation of $J_m(q^2)$; 
when added to the term $(a_2)$, which is also assigned (in its entirety) to the 
$g_{\mu\nu}$ part of that same equation, it finally gives rise, by virtue of \1eq{seagull},   
to a contribution that is free of quadratic divergences and   
vanishes in the $q^2=0$ limit, as it should. 
On the other hand, the second term on the rhs of \1eq{I0mod} is  allotted to the mass equation. 
Thus, 
unlike $B(q^2)$,  which unambiguously contributes to the  $q_{\mu} q_{\nu} /q^{2}$ part of the equation for $J_m(q^2)$
[and satisfies automatically  $B(0) =0$], the $A(q^2)$  contributes to 
the  $g_{\mu\nu}$ component of {\it both} equations. 

Let us finally point out that the purely longitudinal 
nature of the vertices $\widetilde V$ (and $V$)
expressed through conditions such as \1eq{totlon},
combined with the fact that we work in the Landau gauge, allows one to obtain all relevant 
contributions simply from the knowledge of the WI (or STI) that these vertices satisfy,
without the need to construct them {\it explicitly}. 
This is particularly important in the case of the vertex $\widetilde{V}_{\nu\rho\sigma\tau}$, 
appearing in graph $(a_5)$; indeed, constructing this vertex explicitly
would constitute an arduous task, given the complicated STIs that it satisfies  
when contracted by the momentum of any of its three quantum legs.

\section{\label{oneloop}The ``one-loop dressed'' mass equation: Concise derivation}

\begin{figure}
\includegraphics[scale=.7]{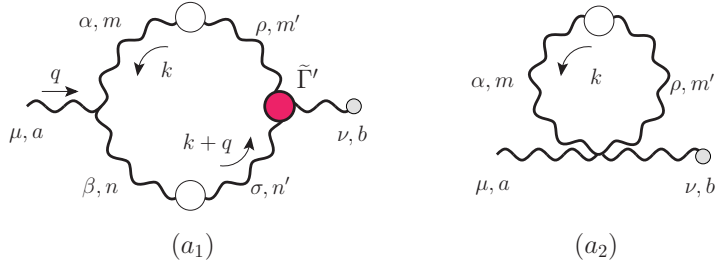} 
\caption{\label{a1-a2} (color online). The {\it one-loop dressed} part of the SDE that contains only gluons. 
Thick lines represent, as previously exlpained, gluon propagators endowed with a momentum-dependent mass. The fully dressed {\it primed} vertex, $\gfullb'$,
enforces gauge invariance in the presence of such a mass.  
The symmetry factor is $1/2$ in both cases. We also show for the reader's convenience (in this and the next figures)  the color and Lorentz indices, as well as  the momentum routing used in our calculations.}
\end{figure}

According to the methodology outlined in the previous section, the one-loop dressed
contribution to the gluon mass equation stems solely from the $\NP$-part of graph $(a_1^{\prime})_{\mu\nu}$, 
to be denoted by $(a^{\s {\widetilde V}}_{1})_{\mu\nu}$.

The first simplification 
stemming from the use of the SDE for the $QB$ propagator (as opposed to the $BB$ employed in~\cite{Aguilar:2011ux}) 
is that the  Landau gauge limit $\xi=0$ may be taken directly, 
in the part of the calculation related to the masses. Indeed, the only source of terms proportional to $\xi^{-1}$ 
(which require special care) is the tree-level part of the $BQ^2$ vertex, which only affect the 
equation for $J(q^2)$  (and can be easily dealt with, following the 
procedure explained in~\cite{Aguilar:2008xm}). 
Then (see \fig{a1-a2} for the Lorentz and color indices as well as the momenta routing used in the following calculation) 
\be
(a^{\s {\widetilde V}}_{1})_{\mu\nu} = \frac12\,\gA 
\int_k\! {\Gamma}^{(0)}_{\mu\alpha\beta}(q,k,-k-q)\Delta^{\alpha\rho}(k)
\Delta^{\beta\sigma}(k+q)\NP_{\nu\rho\sigma}(q,k,-k-q)\,,
\label{gengau}
\ee
where ${\Gamma}^{(0)}_{\mu\alpha\beta}$ is the {\it conventional} three-gluon
vertex of the linear covariant gauges,  
\be
{\Gamma}^{(0)}_{\mu\alpha\beta}(q,k,-k-q)= (q-k)_\beta \,g_{\mu\alpha} +
(2k+q)_\mu \, g_{\alpha\beta} - (2q+k)_\alpha \, g_{\beta\mu},
\label{3g-conv}
\ee
and $\Delta_{\alpha\beta}(k)$ is the totally transverse Landau gauge propagator, namely 
\be
\Delta_{\alpha\beta}(k) = P_{\alpha\beta}(k) \Delta(k^2),
\label{tranprop}
\ee
[notice the minus sign difference with respect to our general definition~\noeq{prop}]. Finally, the trivial color factor $\delta^{ab}$ has been factored out.

As explained in detail in Section~\ref{remind}, 
gauge invariance requires that, when contracted by the momentum of the background leg,   
the vertex $\NP_{\nu\rho\sigma}(q,k,-k-q)$ satisfies the WI of \noeq{winp} with $r=k$ and $p=-(k+q)$, namely 
\be
q^\nu \NP_{\nu\rho\sigma}(q,k,-k-q)= m^2(k)P_{\rho\sigma}(k) - m^2(k+q)P_{\rho\sigma}(k+q).
\label{winp1}
\ee
Now, it is relatively straightforward to recognize that $(a^{\s {\widetilde V}}_{i})_{\mu\nu}$ 
is  proportional to $q_{\mu} q_{\nu}/q^2$ only.
Indeed, the condition of complete longitudinality of $\NP$, given in \1eq{totlon}, becomes 
\be 
P^{\nu\nu'}(q)P^{\alpha\rho}(k)P^{\beta\sigma}(k+q)\NP_{\nu'\rho\sigma}(q,k,-k-q)=0,
\label{transV}
\ee
from which follows immediately that 
\be 
P^{\alpha\rho}(k)P^{\beta\sigma}(k+q)\NP^{\nu}_{\rho\sigma}(q,k,-k-q) = 
\frac{q^{\nu}}{q^2} \left[q^{\nu'}\NP_{\nu'\rho\sigma}(q,k,-k-q)\right]P^{\alpha\rho}(k)P^{\beta\sigma}(k+q).
\label{sma}
\ee
Thus, interestingly enough, the rhs of \1eq{sma} is completely determined from the WI of \1eq{winp}; specifically, using~\noeq{winp1}, we get
\be 
P^{\alpha\rho}(k)P^{\beta\sigma}(k+q)\NP^{\nu}_{\rho\sigma}(q,k,-k-q) = \frac{q^{\nu}}{q^2}
\left[m^2(k) - m^2(k+q)\right] P^{\alpha\rho}(k) P^{\beta}_{\rho}(k+q).
\label{smawi}
\ee
Then, using the elementary tree-level WI
\be 
q^{\mu}\gtree_{\mu\alpha\beta}(q,k,-k-q) = (k+q)^2 P_{\alpha\beta}(k+q) - k^2 P_{\alpha\beta}(k),
\ee
one can show,  after appropriate shifts of the integration variables [\ie $(k+q)\to k$], that indeed
\be
(a^{\s {\widetilde V}}_{1})_{\mu\nu}=\frac{q_\mu q_\nu}{q^2}\frac{\gA}{q^2} \int_k\! m^2(k^2)\left[(k+q)^2-k^2\right] \Delta^{\alpha\rho}(k)\Delta_{\alpha\rho}(k+q),
\label{1l-dr}
\ee
yielding
\be
a^{\s {\widetilde V}}_{1}(q^2) = \frac{\gA }{q^2}
\int_k\! m^2(k^2)\left[(k+q)^2-k^2\right] \Delta^{\alpha\rho}(k)\Delta_{\alpha\rho}(k+q).
\label{1l-dr-1}
\ee

Thus, the one-loop dressed mass equation becomes
\bea
m^2(q^2)&=&\frac{i\gA}{1+G(q^2)} a^{\s {\widetilde V}}_{1}(q^2) 
\nonumber \\
&=&\frac{i\gA}{1+G(q^2)}\frac{1}{q^2}\int_k\!m^2(k^2)\left[(k+q)^2-k^2\right]\Delta^{\alpha\rho}(k)\Delta_{\alpha\rho}(k+q).
\label{new-eq}
\eea

Notice that at the one-loop dressed level the ghost diagrams $(a_3)$ and $(a_4)$ of \fig{glSDE} should also be considered. 
However,  their treatment can be simplified by appealing to some basic properties of the ghost propagator in the Landau gauge, 
established through detailed large-volume lattice simulations, as well as 
a variety of analytic studies~\cite{Boucaud:2008ji,RodriguezQuintero:2010ss,Dudal:2008sp}.
Specifically, we will take for granted  
that the ghost propagator $D$ in the Landau gauge remains massless, $D^{-1}(0)=0$,
while its dressing function $F$ 
is infrared finite, $F(0) =c>0$.

The main implications of these properties for the case at hand is that 
the corresponding fully dressed ghost vertex 
appearing in graph $(a_3)$
does {\it not} need to be modified by the presence of $V$-type vertices.
Specifically, in the absence of a gluon mass  
the vertex $B{\bar c}c$ appearing in $(a_3)$ satisfies 
\bea
q^{\mu}{\Gamma}_{\mu} &=& iD^{-1}(k+q) - iD^{-1}(k)
\nonumber\\ &=& (k+q)^2 F^{-1}(k+q) - k^2 F^{-1}(k).
\label{ghver}
\eea
If $D(q)$ remains massless, as we assume, the only effect of the gluon mass 
is to make the dressing function infrared finite, \ie implement in (\ref{ghver}) the 
replacement $F(q)\to F_m(q)$. Thus, for instance, if $F(q) \sim \ln q^2$, the gluon mass 
induces the qualitative change of the type  $F_m(q) \sim \ln (q^2+m^2)$, accounting for the 
aforementioned infrared finiteness of the ghost dressing function. The point to 
realize is that this proceeds without the need to modify ${\g}_{\mu}$ explicitly, 
by adding to it a $V$-type vertex; ${\g}_{\mu}$ will change only through its implicit 
dependence on the gluon propagators (as well as all other vertices), 
contained inside the diagrams defining its own SDE, 
which have now become ``massive''. 
Given these considerations, the result given in  \1eq{new-eq} exhausts the one-loop dressed case.

Returning now to  \1eq{new-eq}, we emphasize that this equation differs from the corresponding equation derived in~\cite{Aguilar:2011ux}, due to a 
technical subtlety explained in what follows.
Specifically, the method followed in~\cite{Aguilar:2011ux} consisted in taking the trace of 
both sides of \1eq{sdem}; this, in itself, is totally legitimate, given the explicit transversality of both sides,
but, in light of the comments presented in Section~\ref{genmet}, has the disadvantage of 
mixing the  $g_{\mu\nu}$ and the $q_{\mu} q_{\nu} /q^{2}$ components of all contributions. 
As a result, the parts contributing to $J_m(q^2)$ and $m^2(q^2)$ become interwoven. 
This, in turn, forces one to 
resort to an argument analogous to that employed in Section~\ref{genmet}, namely the judicious 
``completion'' of the seagull identity, and the subsequent  
splitting of $A(0)$ between $J_m(q^2)$ and $m^2(q^2)$, but now for nonvanishing $q^2$. 
Specifically, one has that 
\be
f(k,q) = -\frac{(k+q)^2 J_m(k+q) - k^2 J_m(k)}{(k+q)^2 -  k^2}\Delta(k)\Delta(k+q),
\ee
which, of course, in the limit $q^2 \to 0$ reduces to the expression given in \1eq{fk}. 
At this point, one must add and subtract an appropriate term 
that would finally, in the limit $q^2 \to 0$, trigger again the seagull identity.
The term employed in~\cite{Aguilar:2011ux} in order to accomplish this rearrangement was 
\be
\frac{m^2(k+q) - m^2 (k)}{(k+q)^2 -  k^2};
\label{wch}
\ee
instead, the correct term to add and subtract is 
\be
\frac{m^2(k+q) - m^2 (k)}{q^2}.
\label{cch}
\ee
It is relatively straightforward to verify that 
both choices lead to the same limit as \mbox{$q^2 \to 0$}, 
but are obviously different for nonvanishing momentum.
In fact, after restoring the integration sign and all remaining factors, 
\1eq{cch} furnishes precisely the  $g_{\mu\nu}$ part of $P_{\mu\nu}(q) m^2(q^2)$, exactly as  
expected from transversality, while \1eq{wch} clearly does not.    

After this digression, let us return to \1eq{new-eq}. 
The transition to the Euclidean space proceeds by 
using the standard formulas that allow the conversion of the various 
Green's functions from the Minkowski momentum $q^2$ 
to the Euclidean $q^2_{\s{\mathrm{E}}} = -q^2>0$; specifically
\be
\Delta_\mathrm{\s E}(q^2_\mathrm{\s E})=-\Delta(-q^2_\mathrm{\s E}); \qquad m^2_\mathrm{\s E}(q^2_\mathrm{\s E})=  m^2(-q^2_\mathrm{\s E});
\qquad G_\mathrm{\s E}(q^2_\mathrm{\s E})= G(-q^2_\mathrm{\s E});\qquad
\int_k=i\int_{k_\E}.
\label{euc1}
\ee
Then dropping the subscript ``E'', we arrive at the final result
\be
m^2(q^2)=-\frac{\gA}{1+G(q^2)}\frac{1}{q^2}\int_k\!m^2(k^2)\Delta^{\alpha\rho}(k)\Delta_{\alpha\rho}(k+q)\left[(k+q)^2-k^2\right].
\label{1l-meq}
\ee

Consider finally the limit of the above equation as $q^2\to 0$. 
Using the general Taylor expansion [with $y=k^2$ and $w=(k+q)^2$]
\be
f(w)=f(y)+(w-y)f'(y)+{\cal O}(q^2),
\label{exp}
\ee
(the ``prime'' denotes differentiation with respect to $y$, {\it i.e.},  $f'(y) = df(y)/dy$), 
employing ~\2eqs{funrel}{costheta-rel}, and 
the fact that, in $d=4$, $L(0)=0$~\cite{Aguilar:2009pp},  we obtain
\be
m^2(0)=\frac32g^2C_AF(0)\int_k\!k^2\Delta^2(k^2)[m^2(k^2)]'.
\label{m0}
\ee
Note that the result obtained for this limiting case coincides with that found in~\cite{Aguilar:2011ux}.

\section{\label{twoloop}The ``two-loop dressed'' contributions}

\begin{figure}
\includegraphics[scale=.7]{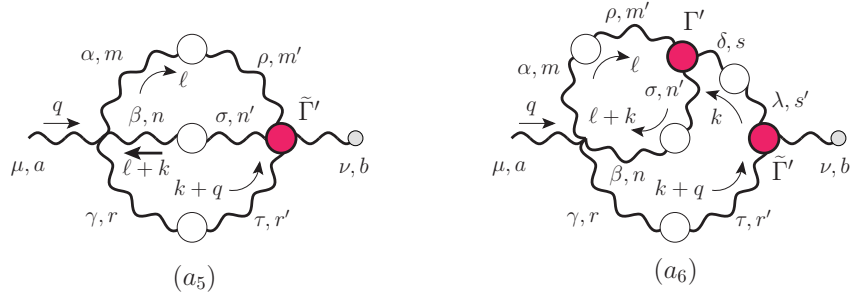} 
\caption{\label{a5-a6}(color online). The {\it two-loop dressed} diagrams. 
The symmetry factors are $1/6$ and $1/2$ respectively.}
\end{figure}

In this section we will study in detail the two-loop dressed diagrams $(a_5)$ and $(a_6)$, 
and their respective contribution to the mass equation (see \fig{a5-a6} for the color and 
Lorentz indices as well as for the momentum routing).
Note that, 
in the alternative $QB$ version of the SDE equation  for the gluon propagator that we consider here,
there are no additional ``two-loop dressed'' diagrams. In particular, there are no diagrams 
involving the fully dressed $BQ{\bar c} c$ vertices (the fourths subset in the usual $BB$ version of the SDE), 
simply because these vertices  cannot be joined 
in any way with the conventional tree-level vertices appearing on the other side of the  
(would be) diagram, where the $Q$-type gluon enters.

\subsection{General considerations regarding the graph $(a_5)$}

Before switching on the Schwinger mechanism, the graph $(a_5)$,  for an arbitrary value of the gauge-fixing parameter $\xi$, is given by 
\be
(a_5)^{ab}_{\mu\nu}=-\frac{i}6{\Gamma}^{(0)amnr}_{\mu\alpha\beta\gamma}\int_k\!\int_\ell\!\Delta^{\alpha\rho}(\ell)\Delta^{\beta\sigma}(\ell+k)\Delta^{\gamma\tau}(k+q)\gfullb_{\nu\tau\sigma\rho}^{brnm}(-q,k+q,-\ell-k,\ell),
\label{a5gen}
\ee
where the tree-level value of the conventional ($Q^4$) four-gluon vertex is given by
\begin{eqnarray}
{\Gamma}_{\mu\alpha\beta\gamma}^{(0)amnr} &=& -ig^2[f^{arx}f^{xnm}(g_{\mu\beta}g_{\alpha\gamma}-g_{\mu\alpha}g_{\beta\gamma}) + f^{amx}f^{xrn}(g_{\mu\gamma}g_{\alpha\beta}-g_{\mu\beta}g_{\alpha\gamma}) \nonumber \\
&+& f^{anx}f^{xrm}(g_{\mu\gamma}g_{\alpha\beta}-g_{\mu\alpha}g_{\beta\gamma})].
\label{tree-levelfour}
\end{eqnarray}
The fully dressed vertex $BQ^3$ satisfies the following WI  (all momenta entering)
\bea
q^\alpha \gfullb^{abcd}_{\alpha\mu\nu\rho}(q,r,p,t) &=& 
ig^2\left[f^{abx}f^{xcd}\Gamma_{\nu\rho\mu}(p,t,q+r) + f^{acx}f^{xdb}\Gamma_{\rho\mu\nu}(t,r,q+p) \right.\nonumber\\
&+&\left. f^{adx}f^{xbc}\Gamma_{\mu\nu\rho}(r,p,q+t)\right],
\label{WI-4g}
\eea
where it should be emphasized that on the rhs appear the conventional (and not the BFM) trilinear vertices,   
which satisfy STIs with respect to all their legs, \ie 
\be
q^\alpha \Gamma_{\alpha\mu\nu}(q,r,p) =  F(q)[ p^2 J(p^2)P^\alpha_\nu(p)H_{\alpha\mu}(p,q,r) - r^2 J(r^2)P^\alpha_\mu(r)H_{\alpha\nu}(r,q,p)], 
\label{STI3g}
\ee
and cyclic permutations~\cite{Ball:1980ax}.

Let us next switch on the  Schwinger mechanism. Then, 
both sides of the WI of \1eq{WI-4g} must be replaced by ``primed'' vertices. Now, the 
``primed'' three-gluon vertices appearing on the rhs are of the type $Q^3$, namely 
$\Gamma^{\prime}_{\alpha\mu\nu}(q,r,p) = \Gamma_{m \alpha\mu\nu}(q,r,p) + {V}_{\alpha\mu\nu}(q,r,p)$, where $\Gamma_{m}$ 
satisfies \1eq{STI3g} with $J \to J_m$ (and $H\to H_m$, which, however, we refrain from indicating), while $V$ must satisfy, correspondingly, 
\be
q^\alpha V_{\alpha\mu\nu}(q,r,p) = F(q)[m^2(r^2)P^\alpha_\mu(r)H_{\alpha\nu}(r,q,p)-m^2(p^2)P^\alpha_\nu(p)H_{\alpha\mu}(p,q,r)] ,
\label{STI-V-3g}
\ee
and cyclic permutations.
Note the difference between \1eq{STI-V-3g} and the corresponding relation satisfied 
by ${\widetilde V}$, given in \1eq{winp}: the latter is an Abelian WI with no reference to the 
ghost sector, while the former is an STI, depending explicitly on the ghost-related quantities 
$F$ and $H$. 

Then, the only possibility for maintaining the original WI of \1eq{WI-4g} intact  is if the quadrilinear  vertex on its lhs    
gets also modified into a vertex satisfying the identity
\bea
q^\alpha \NV^{abcd}_{\alpha\mu\nu\rho}(q,r,p,t) &=&q^\alpha \left[
\gfullb^{abcd}_{{m}\,\alpha\mu\nu\rho}(q,r,p,t)+\Vbqq^{abcd}_{\alpha\mu\nu\rho}(q,r,p,t)\right]\nonumber \\
&=& ig^2 \left[f^{abx}f^{xcd}\qqq'_{\nu\rho\mu}(p,t,q+r) + f^{acx}f^{xdb}\qqq'_{\rho\mu\nu}(t,r,q+p) \right.\nonumber\\
&+&\left. f^{adx}f^{xbc}\qqq'_{\mu\nu\rho}(r,p,q+t)\right],
\label{bqqq-WI}
\eea
where $\bqqm$ and $\Vbqq$ satisfy separately the no-pole ($\qqqm$) and pole ($\Vqqq$) part of the trilinear $Q^3$ vertex, respectively.
In particular, the $\Vbqq$ part, which we will be of central importance in what follows, satisfies
\bea
q^\alpha \Vbqq^{abcd}_{\alpha\mu\nu\rho}(q,r,p,t) &=& 
ig^2\left[f^{abx}f^{xcd}V_{\nu\rho\mu}(p,t,q+r) + f^{acx}f^{xdb}V_{\rho\mu\nu}(t,r,q+p) \right.\nonumber\\
&+&\left. f^{adx}f^{xbc}V_{\mu\nu\rho}(r,p,q+t)\right]. 
\label{WI-V-4g}
\eea

\subsection{The contribution $a_5^{\widetilde{V}}(q^2)$.}

Let us now focus on the part of the diagram $(a_5)$ that contains the pole component $\widetilde{V}$ 
of the fully-dressed $BQ^3$ vertex $\NV$, to be denoted 
by $(a_5^{\widetilde{V}})$. 
The projection to the Landau gauge is straightforward, since  there are no $\xi^{-1}$ terms anywhere in this diagram, 
and  one obtains
\begin{equation}\label{diagram1}
(a_5^{\widetilde{V}})_{\mu\nu}^{ab}=\frac{i}{6}{\Gamma}_{\mu\alpha\beta\gamma}^{(0)amnr}
\int_k\int_\ell\Delta^{\alpha\rho}(\ell)\Delta^{\beta\sigma}(\ell+k)\Delta^{\gamma\tau}(k+q)\widetilde{V}_{\nu\tau\sigma\rho}^{brnm}(-q,k+q,-\ell-k,\ell),
\end{equation}
where all gluon propagators assume the transverse form of \1eq{tranprop}.

We can apply the totally longitudinally coupled condition satisfied by $\widetilde{V}_{\lambda\tau\sigma\rho}$, 
\begin{equation}
P_\nu^\lambda(q)P^{\gamma\tau}(k+q)P^{\beta\sigma}(\ell+k)P^{\alpha\rho}(l)\widetilde{V}_{\lambda\tau\sigma\rho}(-q,k+q,-\ell-k,\ell)=0,
\label{BQQQlongitudinally}
\end{equation}
to write (\ref{diagram1}), after splitting $P_\nu^\lambda(q)$, as follows
\begin{equation}
(a_5^{\widetilde{V}})_{\mu\nu}^{ab}=\frac{i}{6}{\Gamma}_{\mu\alpha\beta\gamma}^{(0)amnr}\frac{q_\nu}{q^2}\int_k\int_\ell\Delta^{\alpha\rho}(\ell)\Delta^{\beta\sigma}(\ell+k)\Delta^{\gamma\tau}(k+q)q^\lambda\widetilde{V}_{\lambda\tau\sigma\rho}^{brnm}(-q,k+q,-\ell-k,\ell).
\label{diagram2}
\end{equation}
Using then the WI~\noeq{WI-4g} adapted to the present kinematics,
as well as the results
\bea
f^{brx}f^{xnm}\Gamma^{(0)amnr}_{\mu\alpha\beta\gamma} &=& i\frac{3}{2}\gA^2\delta^{ab}(g_{\mu\alpha}g_{\beta\gamma}-g_{\mu\beta}g_{\alpha\gamma}); \nonumber \\
f^{bnx}f^{xmr}\Gamma^{(0)amnr}_{\mu\alpha\beta\gamma} &=& i\frac{3}{2}\gA^2\delta^{ab}(g_{\mu\gamma}g_{\alpha\beta}-g_{\mu\alpha}g_{\beta\gamma}); \nonumber \\
f^{bmx}f^{xrn}\Gamma^{(0)amnr}_{\mu\alpha\beta\gamma} &=& i\frac{3}{2}\gA^2\delta^{ab}(g_{\mu\beta}g_{\alpha\gamma}-g_{\mu\gamma}g_{\alpha\beta});
\label{prefactors}
\eea
one finds that each one of the terms of the WI gives rise to an integral of the form
\begin{equation}
t_{j}^{\mu}(q)=t_j(q^2)q^{\mu};\quad j=1,2,3
\label{Bterms}
\end{equation}
so that~\1eq{diagram2} can be written as
\begin{equation}
(a_5^{\widetilde{V}})_{\mu\nu}^{ab}=\frac{i}{4}\gAsq\delta^{ab}\frac{q_\mu q_\nu}{q^2}\sum_{j=1}^3 t_j(q^2),
\label{diagram3}
\end{equation}
with
\begin{eqnarray}
t_1(q^2)&=&\frac{1}{q^2}(q_\beta g_{\alpha\gamma}-q_\alpha g_{\beta\gamma})\int_k\int_\ell\Delta^{\alpha\rho}(\ell)\Delta^{\beta\sigma}(\ell+k)\Delta^{\gamma\tau}(k+q)V_{\sigma\rho\tau}(-\ell-k,\ell,k); \nonumber \\
t_2(q^2)&=&\frac{1}{q^2}(q_\alpha g_{\beta\gamma}-q_\gamma g_{\alpha\beta})\int_k\int_\ell\Delta^{\alpha\rho}(\ell)\Delta^{\beta\sigma}(\ell+k)\Delta^{\gamma\tau}(k+q)V_{\rho\tau\sigma}(\ell,k+q,-q-\ell-k); \nonumber \\
t_3(q^2)&=&\frac{1}{q^2}(q_\gamma g_{\alpha\beta}-q_\beta g_{\alpha\gamma})\int_k\int_\ell\Delta^{\alpha\rho}(\ell)\Delta^{\beta\sigma}(\ell+k)\Delta^{\gamma\tau}(k+q)V_{\tau\sigma\rho}(k+q,-\ell-k,-q+\ell). \nonumber \\
\label{integra1}
\end{eqnarray}

Now it turns out that, after the appropriate shifts in the momenta, relabeling the Lorentz dummy indices, and applying the Bose symmetry of $V_{\alpha\beta\gamma}$, the three terms are actually equal. Then,  using the totally longitudinally coupled condition for the  vertex $V$, 
and the fact that, in the Landau gauge, $k_\tau\Delta^{\gamma\tau}(k+q)=-q_\tau\Delta^{\gamma\tau}(k+q)$, we get
\bea
t(q^2) &=& \sum_{j=1}^3 t_j(q^2)\nonumber \\
&=& 3\frac{q_\tau}{q^2}(q_\alpha g_{\beta\gamma}-q_\beta g_{\alpha\gamma})\int_k \Delta^{\gamma\tau}(k+q)
\int_\ell \Delta^{\alpha\rho}(\ell)\Delta^{\beta\sigma}(\ell+k)\frac{k^\lambda}{k^2} V_{\sigma\rho\lambda}(-\ell-k,\ell,k). \nonumber \\
\label{B1}
\eea
Then, the integral over $\ell$ is a function of $k$ (but not of $q$), and has two free Lorentz indices, $\alpha$ and $\beta$, so that 
\be
\int_\ell \Delta^{\alpha\rho}(\ell)\Delta^{\beta\sigma}(\ell+k)\frac{k^\lambda}{k^2} V_{\sigma\rho\lambda}(-\ell-k,\ell,k) = A(k^2)g^{\alpha\beta}+B(k^2)k^\alpha k^\beta. 
\label{B1a}
\ee
Therefore, 
\be
t(q^2) = 3\frac{q_\tau}{q^2}(q_\alpha g_{\beta\gamma}-q_\beta g_{\alpha\gamma})\int_k \Delta^{\gamma\tau}(k+q) 
\,[A(k^2)g^{\alpha\beta}+B(k^2)k^\alpha k^\beta].
\label{B1b}
\ee
But this term vanishes, regardless of the closed form of $A(k^2)$ and $B(k^2)$, 
because the prefactor is antisymmetric under 
the exchange $\alpha\leftrightarrow\beta$, whereas the integral is symmetric. 

Thus, one finally arrives at the important result 
\be
a_5^{\widetilde{V}}(q^2) =0,
\label{surprise}
\ee
namely that the graph $(a_5)$ makes no contribution to the gluon mass equation. 

\subsection{The contributions from graph $(a_6)$}

Let us now consider the graph $(a_6)$,  for a general value of the gauge-fixing parameter $\xi$. 
This graph contains the $BQ^2$ and $Q^3$  fully-dressed three-gluon vertices, 
namely $\gfullb$ and $\Gamma$. 
We proceed directly to the massive situation,   
where these two vertices have been replaced by their ``primed'' counterparts, namely 
\bea
(a_6)^{ab}_{\mu\nu} &=& \frac34 i \gAsq\delta^{ab} \left(g_{\mu\alpha}g_{\beta\gamma}-g_{\mu\beta}g_{\alpha\gamma}\right)
\int_k\! \Delta^{\gamma\tau}(k+q)\Delta^{\delta\lambda}(k) \gfullb^{\prime}_{\nu\tau\lambda}(-q,k+q,-k)
\nonumber \\
&\times&
\int_\ell\!\Delta^{\alpha\rho}(\ell)\Delta^{\beta\sigma}(\ell+k)\Gamma^{\prime}_{\sigma\rho\delta}(-\ell-k,\ell,k). 
\label{a6}
\eea
Evidently, $\gfullb^{\prime}$ and $\Gamma^{\prime}$
contain the pole parts $\widetilde V$ and $V$, respectively, satisfying the general 
properties mentioned earlier. 
We will next isolate the terms proportional to  $\widetilde V$ and $V$, 
since it is these terms that determine the 
corresponding contribution of the entire graph $(a_6)$ to the gluon mass equation.  
This amounts to writing the product $\gfullb^{\prime}\Gamma^{\prime}$ as 
\bea
\gfullb^{\prime}\Gamma^{\prime} &=& (\gfullb_m + {\widetilde V}) (\Gamma_m + V)
\nonumber \\
&=& \gfullb_m \Gamma_m + {\widetilde V} \Gamma_m  + \gfullb_m  V + {\widetilde V} V,  
\label{Lpt}
\eea
and considering only the last three terms. 

\subsubsection{\label{van}Vanishing of the terms proportional to $V$}

To proceed with the demonstration, note that if we were in the Landau gauge, \ie if the gluon propagators in  \1eq{a6} 
had the fully transverse form of \1eq{tranprop},
then $V$ would vanish identically, due to its property of complete longitudinality, given that it is an internal vertex ($Q^3$-type). 
The limit $\xi=0$ may be taken directly in the part of the graph involving the term ${\widetilde V} V$, and therefore 
this term vanishes immediately. As for the combination  $\widetilde{\qqq}_{m} V$,  
one can take directly the 
limit $\xi=0$ everywhere, thus making it vanish, 
except for the term that contains the tree-level part of $\widetilde{\qqq}_{m}$ that is proportional to 
$\xi^{-1}$.
Specifically, the tree-level part of the $BQ^2$ vertex is given by  
\be
\gtreeb_{\nu\tau\lambda}(-q,k+q,-k) =
\gtree_{\nu\tau\lambda}(-q,k+q,-k) - \xi^{-1} \Gamma^{\s{\rm P}}_{\nu\tau\lambda}(-q,k+q,-k),
\label{PTdec}
\ee 
where the  purely longitudinal ``pinch part'' $\Gamma^{\s{\rm P}}$ is given by
\be
\Gamma^{\s{\rm P}}_{\nu\tau\lambda}(-q,k+q,-k) = - g_{\nu\lambda} (k+q)_{\tau} - g_{\tau\nu} k_{\lambda}.
\label{PTvert}
\ee
The contraction of this term with the propagators $\Delta^{\gamma\tau}(k+q)\Delta^{\delta\lambda}(k)$ (with $\xi$ still general)
yields
\be
\xi^{-1} \Gamma^{\s{\rm P}}_{\nu\tau\lambda}(-q,k+q,-k)\Delta^{\gamma\tau}(k+q)\Delta^{\delta\lambda}(k) 
= \frac{k^\delta}{k^2} \Delta_{\nu}^\gamma(k+q)+\frac{(k+q)^\gamma}{(k+q)^2} \Delta_\nu^\delta(k).
\label{ptcan}
\ee
In this way, the $\xi^{-1}$ term cancels, making the limit $\xi \to 0$ smooth. 
So, after setting $\xi = 0$, the second term in \1eq{ptcan} vanishes, because $V$ will be contracted by three 
transverse projectors; 
on the other hand, the first term survives, since $V$ is contracted only by two. 
However, as we will see now, this last term finally also vanishes, due to a different reason. 
Specifically, denoting this contribution by $(a_6^V)_{\mu\nu}$ (thus factoring out the trivial color structure $\delta^{ab}$), we have 
\be
(a_6^V)_{\mu\nu} =\frac34i\gAsq\left(g_{\mu\alpha}g_{\beta\gamma}-g_{\mu\beta}g_{\alpha\gamma}\right)
\int_k\! \Delta_{\nu}^\gamma(k+q) \int_\ell\!\Delta^{\alpha\rho}(\ell)\Delta^{\beta\sigma}(\ell+k) 
\frac{k^\delta}{k^2} V_{\sigma\rho\delta}(-\ell-k,\ell,k).
\label{a6V}
\ee
Now, the integral $\int_\ell$ contains $k$ but no $q$, and has two free Lorentz indices, $\alpha$ and $\beta$; therefore,
it can only be proportional to $A(k^2) g_{\alpha\beta}$ and $B(k^2) k_{\alpha}k_{\beta}$. But, since both these terms are symmetric 
under $\alpha \leftrightarrow\beta$, while the prefactor is antisymmetric, this term vanishes.

\subsubsection{The term $a_6^{\widetilde V}(q^2)$}

Let us finally consider the term ${\widetilde V} \Gamma_m$ in \1eq{Lpt}, to be denoted by $(a_6^{\widetilde V})_{\mu\nu}$. 
It is convenient to define the quantity 
\be
\Y_{\delta}^{\alpha\beta}(k)=\int_\ell\!\Delta^{\alpha\rho}(\ell)\Delta^{\beta\sigma}(\ell+k)\Gamma_{\sigma\rho\delta}(-\ell-k,\ell,k),
\label{defY}
\ee
corresponding to the subdiagram on the upper left corner of $(a_6)$.
Then, $(a_6^{\widetilde V})_{\mu\nu}$ is given by
\be
(a_6^{\widetilde V})_{\mu\nu} = \frac34i\gAsq\left(g_{\mu\alpha}g_{\beta\gamma}-g_{\mu\beta}g_{\alpha\gamma}\right)
\int_k\! \Y_{\delta}^{\alpha\beta}(k) \Delta^{\gamma\tau}(k+q)\Delta^{\delta\lambda}(k) {\widetilde V}_{\nu\tau\lambda}(-q,k+q,-k)
\ee
Then, using once again \3eqs{winp}{transV}{sma}, we obtain
\bea
(a_6^{\widetilde V})_{\mu\nu}
&=&\frac34i\gAsq\left(g_{\mu\alpha}g_{\beta\gamma}-g_{\mu\beta}g_{\alpha\gamma}\right)\frac{q^\nu}{q^2}\int_k\!\left[m^2(k)-m^2(k+q)\right]\Delta^{\delta}_{\lambda}(k)\Delta^{\gamma\lambda}(k+q)\Y_{\delta}^{\alpha\beta}(k) \nonumber \\
&=&\frac{q_\mu q_\nu}{q^2} a_6^{\widetilde V}(q^2)\,,
\eea
and so
\be
a_6^{\widetilde V}(q^2)=\frac34i\gAsq\left(q_{\alpha}g_{\beta\gamma}-q_{\beta}g_{\alpha\gamma}\right)\frac1{q^2}\int_k\!\left[m^2(k)-m^2(k+q)\right]\Delta^\delta_\lambda(k)\Delta^{\gamma\lambda}(k+q)\Y_{\delta}^{\alpha\beta}(k).
\label{fnal}
\ee

At this point it is easy to show that the integral $Y$ is antisymmetric under the $\alpha\leftrightarrow\beta$ exchange; thus, given also the antisymmetry of the $a_6^{\widetilde V}$ prefactor under the same exchange, one can write 
\be
\Y^{\alpha\beta}_{\delta}(k)=(k^\alpha g^\beta_\delta-k^\beta g^\alpha_\delta)\Y(k^2);\qquad \Y(k^2)=\frac1{d-1}\frac{1}{k^2}\,k_\alpha g_\beta^\delta \Y_{\delta}^{\alpha\beta}(k)\,,
\label{Ydef}
\ee 
which gives us the final result
\bea
a_6^{\widetilde V}(q^2) &=&\frac3{4}i\frac{\gAsq}{q^2}\int_k\!
m^2(k^2)[(k+q)^2-k^2][Y(k+q)+Y(k)]\Delta^\delta_\lambda(k)\Delta_\delta^\lambda(k+q)\nonumber \\
&+&\frac3{4}i\frac{\gAsq}{q^2}(q^2g_{\delta\gamma}-2q_\delta q_\gamma)\int_k\!
m^2(k^2)[Y(k+q)-Y(k)]\Delta^\delta_\lambda(k)\Delta^{\gamma\lambda}(k+q).
\label{a6Vtilde} 
\eea

\subsection{Explicit check of the two-loop dressed blockwise transversality}

The vanishing of the term $a_5^{\widetilde V}$ may appear somewhat surprising,
since, in the  PT-BFM framework that we employ, it is exactly the $V$ and
$\widetilde{V}$ type  of vertices that  allow for the appearance  of a
dynamically generated  gluon mass in  a gauge-invariant way.  Thus, one
might  wonder  whether the  result~\noeq{surprise} is in any way  at odds 
with the characteristic property of the blockwise transversality, 
mentioned earlier. 

To show that  this is not the case, let us  contract diagram  $(a_5)$ 
with the physical  momentum $q$; after  carrying  out the
usual      splitting      of      the     $BQ^3$      full      vertex
$\widetilde\Gamma\to\widetilde\Gamma_m+\widetilde    V$,   using   the
result~\noeq{surprise}  and   applying  the  WI~\noeq{WI-4g}   to  the
remaining term, we get
\bea
q^\nu(a_5^{\widetilde{\Gamma}'})_{\mu\nu}^{ab} &=& q^\nu(a_5^{\widetilde{\Gamma}_m})_{\mu\nu}^{ab} \nonumber \\
&=& \frac{3}{4}ig^4C_A^2\delta^{ab}(g_{\mu\beta}g_{\alpha\gamma}-g_{\mu\alpha}g_{\beta\gamma}) \int_k\!\int_\ell\!\Delta^{\alpha}_{\rho}(\ell)\Delta^{\beta}_{\sigma}(\ell+k)\Delta^{\gamma}_{\tau}(k+q)\Gamma_{m}^{\sigma\rho\tau}(-\ell-k,\ell,k).\nonumber \\
\label{4}
\eea

As far  as the contribution of  diagram $(a_6)$ is  concerned, we know
that it cannot be  projected  directly to  the Landau  gauge,
because  the  tree-level  part  of  the  fully-dressed  $BQ^2$  vertex
contains  terms proportional to  $1/\xi$.     
However, proceeding as in
subsection~\ref{van}, one writes
\be
\widetilde{\Gamma}'_{\nu\tau\epsilon}(-q,k+q,-k) = \widetilde{\Gamma}^{\prime\, {\rm reg}}_{\nu\tau\epsilon}(-q,k+q,-k) - \xi^{-1}\Gamma^\s{\rm P}_{\nu\tau\epsilon}(-q,k+q,-k),
\label{6}
\ee
where evidently the regular part $\widetilde\Gamma^{\prime\, {\rm reg}}$ differs from the usual $\widetilde\Gamma'$ by a tree-level term. For the regular term, after using the identity
\be
q^\nu\widetilde{\Gamma}^{\prime\,{\rm reg}}_{\nu\tau\epsilon}(-q,k+q,-k) = \Delta^{-1}(k)P_{\tau\epsilon}(k) - \Delta^{-1}(k+q)P_{\tau\epsilon}(k+q),
\ee
one gets
\bea
q^\nu(a_6^{\widetilde{\Gamma}^{\prime\,{\rm reg}}})_{\mu\nu}^{ab} &=& \frac{3}{4}ig^4C_A^2\delta^{ab}(g_{\mu\alpha}g_{\beta\gamma}-g_{\mu\beta}g_{\alpha\gamma})\int_k\int_\ell[\Delta(k+q)-\Delta(k)]P_{\tau\delta}(k)P^{\gamma\tau}(k+q) \nonumber \\
&\times & \Delta^{\alpha}_{\rho}(\ell)\Delta^{\beta}_{\sigma}(\ell+k)\Gamma_m^{\sigma\rho\delta}(-\ell-k,\ell,k).
\label{9}
\eea

\begin{figure}
\includegraphics[scale=.55]{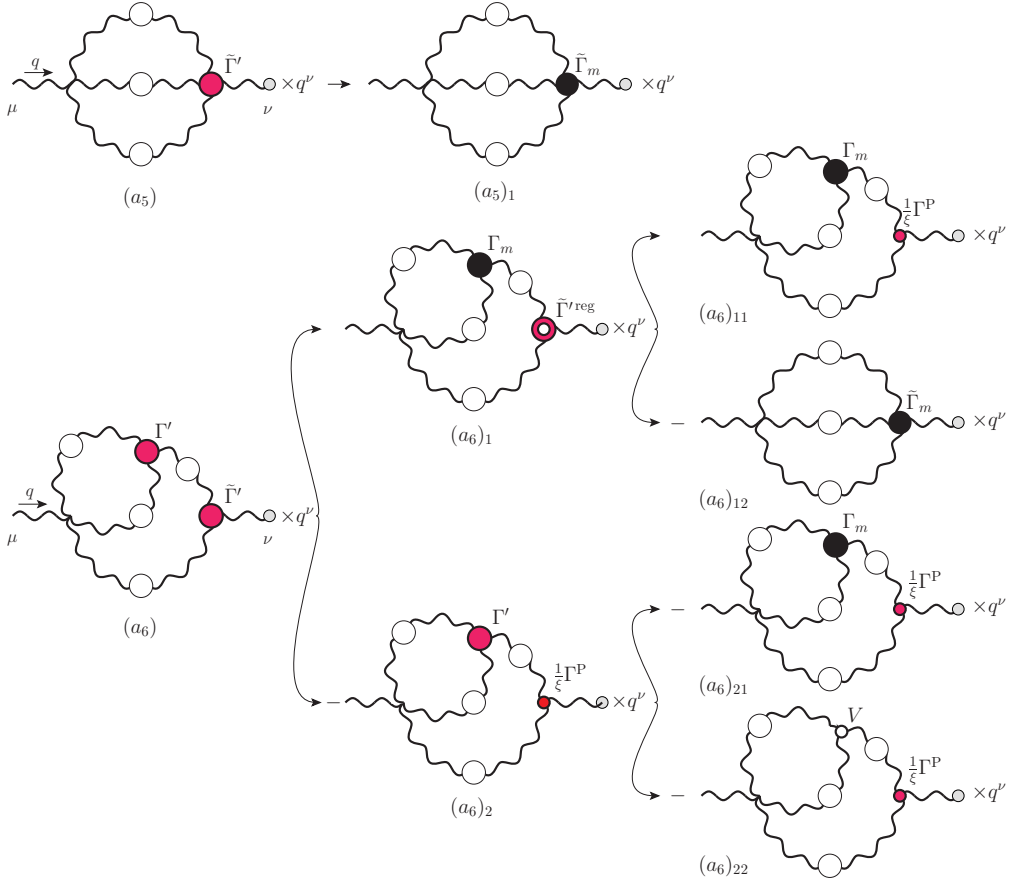}    
\caption{\label{2ldrWI}(color online). Diagrammatic realization of the WI for the two-loop dressed diagrams in the Landau gauge. 
One has the cancellations $(a_5)_1+(a_6)_{12}=0$ and $(a_6)_{11}+(a_6)_{21}=0$, while the   
term $(a_6)_{22}$ vanishes.}
\end{figure}

For the pinch part, after using \1eq{ptcan} to cancel the $\xi^{-1}$ dependence, one obtains (now in the Landau gauge)
\bea
(a_6^{\Gamma^\s{\rm P}})_{\mu\nu}^{ab} &=& \frac{3}{4}ig^4C_A^2\delta^{ab}(g_{\mu\beta}g_{\alpha\gamma}-g_{\mu\alpha}g_{\beta\gamma})\int_k\!\left[\frac{k^\delta}{k^2}\Delta_\nu^\gamma(k+q) + \frac{(k+q)^\gamma}{(k+q)^2}\Delta^\delta_\nu(k)\right] \nonumber \\
&\times & \int_\ell\!\Delta^{\alpha\rho}(\ell)\Delta^{\beta\sigma}(\ell+k)\Gamma'_{\sigma\rho\delta}(-\ell-k,\ell,k).
\label{10}
\eea
On the other hand, observing that
\bea
q^\nu\bigg[\frac{k^\delta}{k^2}\Delta_\nu^\gamma(k+q) + \frac{(k+q)^\gamma}{(k+q)^2}\Delta^\delta_\nu(k)\bigg] &=& \Delta^{\gamma\delta}(k)-\Delta^{\gamma\delta}(k+q) \nonumber \\
&+& [\Delta(k+q)-\Delta(k)]P^{\delta\nu}(k)P_\nu^\gamma(k+q),
\label{11}
\eea 
\1eq{10}  becomes  
\bea 
q^\nu(a_6^{\Gamma^P})_{\mu\nu}^{ab}  &=&
\frac{3}{4}ig^4C_A^2\delta^{ab}(g_{\mu\beta}g_{\alpha\gamma}-g_{\mu\alpha}g_{\beta\gamma})\left\lbrace\int_k\!\int_\ell\!\Delta^{\gamma}_{\delta}(k)\Delta^{\alpha}_{\rho}(\ell)\Delta^{\beta}_{\sigma}(\ell+k)\Gamma_{
m}^{\sigma\rho\delta}(-\ell-k,\ell,k)\right.    \nonumber    \\    &-&
\int_k\!\int_\ell\!\Delta^{\gamma\delta}(k+q)\Delta^{\alpha\rho}(\ell)\Delta^{\beta\sigma}(\ell+k)\Gamma'_{\sigma\rho\delta}(-\ell-k,\ell,k)
\nonumber                            \\                            &+&
\left.\int_k\int_\ell[\Delta(k+q)-\Delta(k)]P_{\delta}^{\nu}(k)P_\nu^\gamma(k+q)\Delta^{\alpha}_{\rho}(\ell)\Delta^{\beta}_{\sigma}(\ell+k)\Gamma_{m}^{\sigma\rho\delta}(-\ell-k,\ell,k)\right\rbrace. \nonumber
\\  \eea 
Clearly  the  first  term integrates  to  zero, since,  being
independent of $q$, it cannot  saturate its free index $\mu$, while the third
term cancels exactly against~\noeq{9}. As  far as the second  term is
concerned, notice that  it still contains a pole  part, since the total
longitudinality  condition of \1eq{totlon}
cannot be triggered in  this case.  However,
after  splitting  the  full   vertex  $\Gamma'$  one  finds  that  the
$\Gamma_m$ part  cancels with the  term~\noeq{4}, while it is  easy to
show that the pole part $V$  vanishes along the same lines described when dealing 
with graph $a_5^{\widetilde V}$.

The realization of the two-loop dressed blockwise transversality in the Landau gauge is shown diagrammatically in~\fig{2ldrWI}. 

\section{\label{fullmasseq} The full mass equation}

\begin{figure}
\includegraphics[scale=1]{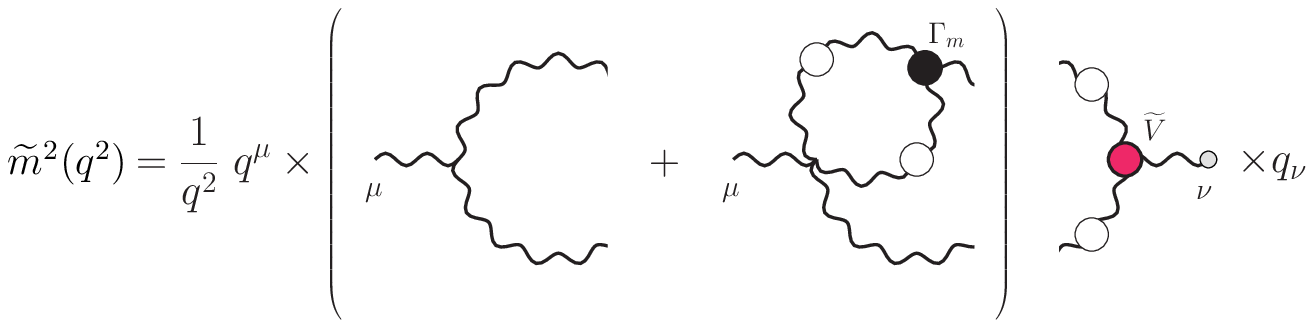} 
\caption{\label{diagrammaticmass}(color online). Diagrammatic representation of the condensed operations 
leading to the all-order gluon mass equation, where we have introduced the shorthand notation   
${\widetilde m}^2 (q^2) = m^2(q^2)[1+G(q^2)]$. All internal propagators are in the Landau gauge.}
\end{figure}

After these rather technical considerations, we are 
now in position to write down the all-order mass equation. 
Using \1eq{masseq}, together with the results~\noeq{1l-dr-1} and~\noeq{a6Vtilde}, one finds
\begin{eqnarray}
m^2(q^2) &=& \frac{i}{1+G(q^2)}\left[a_1^{\widetilde{V}}(q^2) + a_6^{\widetilde{V}}(q^2)\right]\nonumber \\
&=&\frac{ig^2 C_A}{1+G(q^2)}\frac{1}{q^2}\int_k m^2(k^2)[(k+q)^2 - k^2]\Delta^{\alpha\rho}(k)\Delta_{\alpha\rho}(k+q) \nonumber \\
&\times & \left\lbrace 1 + \frac{3}{4}ig^2C_A[Y(k+q) + Y(k)]\right\rbrace \nonumber \\
&-& \frac{3}{4}\frac{g^4 C_A^2}{1 + G(q^2)}\frac{1}{q^2}(q^2 g_{\delta\gamma}-2q_\delta q_\gamma)\int_k m^2(k^2)[Y(k+q)-Y(k)]\Delta_\epsilon^\delta(k)\Delta^{\gamma\epsilon}(k+q).\nonumber \\
\end{eqnarray}
The transition to the Euclidean momenta can be performed by using the standard formulas \noeq{euc1} supplemented with the relation $Y_E(q^2_E)=-iY(-q_E^2)$; then one obtains (suppressing the ``E'' subscript as usual)
\begin{eqnarray}
m^2(q^2) &=& -\frac{g^2 C_A}{1+G(q^2)}\frac{1}{q^2}\int_k m^2(k^2)[(k+q)^2 - k^2]\Delta^{\alpha\rho}(k)\Delta_{\alpha\rho}(k+q) \nonumber \\
&\times & \left\{ 1 - \frac{3}{4}g^2C_A[Y(k+q) + Y(k)]\right\} \nonumber \\
&+& \frac{3}{4}\frac{g^4 C_A^2}{1 + G(q^2)}\frac{1}{q^2}(q^2 g_{\delta\gamma}-2q_\delta q_\gamma)\int_k m^2(k^2)[Y(k+q)-Y(k)]\Delta_\epsilon^\delta(k)\Delta^{\gamma\epsilon}(k+q).\nonumber \\
\label{fullmass} 
\end{eqnarray}
Interestingly enough, the full diagrammatic analysis presented in Sections~\ref{oneloop} and~\ref{twoloop},
in conjunction with the methodology developed in Section~\ref{genmet}, may be pictorially summarized, in a rather concise way, 
as shown in Fig~\ref{diagrammaticmass}.   

Note that  the quantities entering in \1eq{fullmass} are bare, and must eventually undergo renormalization, 
following the general methodology outlined in various earlier works~\cite{Aguilar:2009pp,Aguilar:2009ke}. To be sure, 
the renormalized version of \1eq{fullmass} will involve some of the cutoff-dependent 
renormalization constants $Z_i$ introduced during the renormalization procedure, 
in a way analogous to what happens in the 
case of the integral equation for the dynamical quark mass (see, \eg~\cite{Roberts:1994dr,Aguilar:2010cn}); 
however, for the purposes of the present work, we will ignore these constants 
(replacing them, effectively, by unity, \ie $Z_i\to 1$ ), 
and will simply consider \1eq{fullmass}, assuming that the quantities appearing there are now the renormalized ones.

As has been explained in section ~\ref{genmet}, the mass equation derived in  \1eq{fullmass} 
constitutes one of the two coupled integral equations that govern simultaneously the  
dynamics of $m^2(q^2)$ and $J(q^2)$ [see, for example, \1eq{separ}]. 
If the  corresponding all order integral equation for $J(q^2)$ were known, 
then one could attempt to solve the coupled system, after carrying out the 
additional substitution $\Delta(k^2) = [k^2 J(k^2) + m^2(k^2)]^{-1}$ [{\it viz.} \1eq{massive}]
to all gluon propagators appearing inside the various kernels. 
It turns out that the derivation of the all-order integral equation for $J(q^2)$ is technically far more difficult, 
mainly due to the presence of the fully dressed four-gluon vertex $BQ^3$ [see graph ($a_5$) in Fig.\ref{glSDE}], 
which is a largely unexplored quantity, with a complicated Lorentz and color structure, 
and a vast proliferation of form factors. In fact, 
unlike what happens in the case of the three-gluon vertex $BQ^2$~\cite{Binosi:2011wi}, 
no gauge-technique Ansatz exists for this four-gluon vertex. 
Thus, for the rest of this analysis, we will study \1eq{fullmass} in isolation, 
treating all full propagators appearing in it as external quantities, whose form will be determined 
by resorting to information beyond the SDEs, such as the large-volume lattice simulations. 
Therefore, \1eq{fullmass} is effectively converted into a homogeneous {\it linear} integral equation for 
the unknown function  $m^2(q^2)$. 

Evidently, the quantity $\Y_{\delta}^{\alpha\beta}(k)$ introduced in \1eq{defY} accounts for the 
bulk of the two-loop contribution and depends explicitly on the fully dressed three-gluon vertex $\Gamma$ (of the type $Q^3$), 
in the Landau gauge. 
This Bose-symmetric vertex satisfies the well-known STI~\noeq{STI3g} and its cyclic permutations, which allow, in turn, 
for the reconstruction of the longitudinal 
form factors of $\Gamma$ in terms of $J$, $F$, and the various 
form factors  of the ghost-gluon kernel $H$~\cite{Ball:1980ax}. 
Clearly, the inclusion of the (ten) longitudinal vertex form factors into $\Y_{\delta}^{\alpha\beta}(k)$, 
and through it into \1eq{fullmass}, will give rise to rather complicated expressions, 
whose numerical treatment lies beyond the scope of this work.

In what follows we will simplify this preliminary analysis by considering simply 
the lowest-order perturbative expression for  $\Y$, obtained by substituting tree-level values for all quantities 
appearing in  \1eq{defY}, and using \noeq{Ydef}. It turns out that even so, the resulting 
expression for $\Y$ has sufficient structure to effectively reverse the overall sign of the equation and give rise to 
physically meaningful solutions for $m^2(q^2)$.
In particular, one has
\bea
Y(k^2)&=&\frac1{d-1}\frac{k_\alpha}{k^2}\int_\ell\frac1{\ell^2(\ell+k)^2}P^{\alpha\rho}(\ell)P^{\beta\sigma}(\ell+k)\qqq^{(0)}_{\sigma\rho\beta}(-\ell-k,\ell,k)\nonumber\\
&=&\frac1{d-1}\frac{k_\alpha}{k^2}\left(2g_{\beta\sigma}k_\rho+g_{\sigma\rho}k_\beta-2g_{\beta\rho}k_\sigma\right)\nonumber \\
&\times&\int_\ell\frac{1}{\ell^2(\ell+k)^2}\left[g^{\alpha\rho}g^{\beta\sigma}-\frac1{\ell^2}(\ell^\alpha\ell^\rho g^{\beta\sigma}+g^{\alpha\rho}\ell^\beta\ell^\sigma)+\frac1{\ell^2(\ell+k)^2}\ell^\alpha\ell^\rho(\ell+k)^\beta(\ell+k)^\sigma
\right],\nonumber \\
\eea
and a straightforward calculation yields (in dimensional regularization, Euclidean space)
\be
Y(k^2)=\frac{1}{3(4\pi)^2}\left[\frac{15}4\left(\frac2\epsilon\right)-\frac{15}4\left(\gamma_{\mathrm{E}}-\log4\pi+\log\frac{k^2}{\mu^2}\right)+\frac{63}{12}\right],
\ee
where $\mu$ is the 't Hooft mass introduced at \1eq{dqd}. 
$Y(k^2)$ may be renormalized within the MOM scheme, by simply subtracting its value at $k^2 = {\bar{\mu}}^2$, yielding  
\be
Y_{\mathrm R}(k^2)=-\frac 1{(4\pi)^2}\frac54\log\frac{k^2}{{\bar{\mu}}^2}.
\label{Yappr}
\ee

\section{\label{numan} Numerical analysis}

In this section we carry out a rather thorough numerical analysis of the mass equation derived in the previous sections.

To begin with, let us rewrite the equation (for the $d=4$ case) in a form that will be suited for the ensuing numerical treatment. After setting $x=q^2$, observing that $(k+q)^2=x+y+2\sqrt{xy}\cos\theta$, and using the measure
\be
\int_k = \frac{1}{(2\pi)^3}\int_0^\pi\!\diff \theta\,\sin^{2}\theta\int_0^\infty\! \diff y\, y,
\ee
we obtain
\bea
m^2(x)&=&-\lambda\frac{F(x)}x\int_0^\pi\!\diff\theta\,\sin^2\theta\int_0^\infty\!\diff y\,y\,m^2(y)\Delta(y)\Delta(x+y+2\sqrt{xy}\cos\theta)\nonumber \\
&\times&\left\{A(x,y,\theta) B(x,y,\theta)\left[1-C\left(\Y(x+y+2\sqrt{xy}\cos\theta)+\Y(y)\right)\right]\right.\nonumber \\
&-&\left.CE(x,y,\theta)\left(\Y(x+y+2\sqrt{xy}\cos\theta)-\Y(y)\right)
\right\},
\label{me-gen}
\eea
where we have used the approximation~\cite{Aguilar:2009pp} $1+G(x)\approx F^{-1}(x)$, and set
\bea
A(x,y,\theta)&=&3-\frac{x\sin^2\theta}{x+y+2\sqrt{xy}\cos\theta};\nonumber \\
B(x,y,\theta)&=&x+2\sqrt{xy}\cos\theta;
\nonumber \\
E(x,y,\theta)&=&\frac{xy+x\cos^2\theta(x+2\sqrt{xy}\cos\theta)+2(x+\sqrt{xy}\cos\theta)^2}{x+y+2\sqrt{xy}\cos\theta}.
\eea
and we have defined ($\alpha_s=g^2/4\pi$)
\be
\lambda=\frac{\alpha_s C_A}{2\pi^2},
\label{coupl-lam}
\ee 
In addition, we have introduced the constant $C$, multiplying the contribution to the mass equation that is of pure 
two-loop origin. Of course, the value of $C$ corresponding to the approximate expression \1eq{Yappr} that we employ is fixed, 
namely 
\be
C=3\pi C_A\alpha_s; 
\label{theC}
\ee
however, during a significant part of the ensuing analysis we will treat $C$ as a free parameter.
Thus, essentially, one disentangles $C$ from the value of $\alpha_s$, 
and studies what happens to the gluon-mass equation  
when one varies independently  $\alpha_s$ and $C$.
The reason for doing this is twofold: \n{i} one has the ability to switch off completely the two-loop corrections (by setting $C=0$), and \n{ii}
by varying the value of $C$ one may study, in some additional detail, the quantitative impact of the two-loop contribution. 
Specifically, the philosophy 
underlying point \n{ii} is that, whereas the expression in \1eq{Yappr} furnishes a concrete form for the two-loop correction, by no means does
it exhaust it; thus, by varying $C$ we basically model, in a rather heuristic way,   
further correction that may be added to the ``skeleton'' provided by the $Y(k^2)$ of \1eq{Yappr} (for a fixed value of $\alpha_s$). 
Of course, the case where $C$ admits the actual value of \1eq{theC} will emerge as a special case of this 
general two-parameter study. In view of the ensuing analysis it is convenient to measure $C$ in units of $3\pi C_A$; to this end, we introduce the reduced parameter $C_r= C/3\pi C_A$, and drop the suffix ``$r$'' in what follows. 

The integral equation to solve is a homogenous Fredholm equation of the second kind, and can be rewritten schematically as
\be
m^2(x)=-\lambda\int_0^\pi\!\diff \theta\!\int_0^b\!\diff y\, {\cal K}(x,y,\theta)\,m^2(y),
\ee
where $b=\infty$, but in practice we will choose $b\gg1$ but finite.
A possible way of solving this equation is to expand the unknown function in terms of a suitable function basis, 
and subsequently determine the coefficients of this expansion. In particular~\cite{Press:1992zz,Bloch:1995dd,Bloch:2003yu}, using the  Chebishev polynomials of the first kind $T_k$, one can write 
\be
m^2(x)=\frac{c_0}2+\sum_{k=1}^nc_kT_k(x).
\ee

In order to determine the $n+1$ coefficients characterizing the expansion, one discretizes $x\in[0,b]$, choosing the $x_j$ values that correspond to the extrema of the $n^{\rm th}$ Chebishev polynomial in that interval, \ie
\be
x_j=\frac b2\cos\left(\frac{\pi}{n}(n-j)\right)+\frac b2,\qquad j=0,\dots,n.
\ee
The problem is then reduced to finding the values of $\lambda$ for  
which the matrix $A+\lambda B$ is singular, where 
\be
A_{ij}=\delta T_i\left(\frac{2x_j-b}b\right); \qquad B_{ij}=\delta\int_0^\pi\!\diff \theta\!\int_0^b\!\diff y\, {\cal K}(x_j,y,\theta)\,T_i\left(\frac{2x_j-b}b\right),
\ee
and $\delta=1$, unless $i=0$, in which case it is $1/2$. Specifically one is looking for the smallest positive root $\lambda_s$ of the generalized characteristic polynomial of the matrices $A$ and $-B$. Provided that $\lambda_s$ exists, one can next determine  all the expansion coefficients $c_k$ by simply assigning to $c_0$ a predetermined value (we choose $c_0=1$) and then solving the resulting reduced system; the corresponding value of the coupling constant can then be obtained  through~\1eq{coupl-lam}. The solution can be finally rescaled 
by an arbitrary (positive) constant, 
due to the freedom allowed by the linearity of the equation.

\subsection{The one-loop dressed case}

Let us start our analysis from the one-loop dressed case, which corresponds to setting $C=0$ in~\1eq{me-gen}. 
Specifically, notice that, as $x\to0$,  
this equation reduces to the following nonlinear constraint
\be
m^2(0)=-\frac{3\pi}{4}\lambda\, F(0)\int_0^\infty\!\diff y\,m^2(y){\cal K}_{\rm 1}(y);\qquad {\cal K}_{\rm 1}(y)=\left[{\cal Z}^2(y)\right]',
\label{completem0-1l}
\ee
where ${\cal Z}(y)=y\Delta(y)$ is the gluon dressing function. Note that the constraint of \1eq{completem0-1l} is identical to that 
derived in~\cite{Aguilar:2011ux}; remember, however, that the full one-loop equation for arbitrary momenta 
is different, for the reasons explained in Section~\ref{oneloop}.

As explained in~\cite{Aguilar:2011ux}, the usefulness of \1eq{completem0-1l} lies in the fact that, already at this level, one 
may recognize the difficulty in obtaining   physical solutions (\ie positive definite in the entire momentum range), 
which can be ultimately traced down to the ``wrong'' sign in front of the equation. 
In addition, one may explore  
the conditions that might finally overcome this difficulty, without having to solve the integral equation in its full complexity.
In particular, it is clear from \1eq{completem0-1l} 
that, in order to have a possibility of obtaining physically meaningful solutions, 
the kernel ${\cal K}_1$ must display a ``sufficiently deep'' negative region, which might eventually 
counterbalance the overall minus sign; indeed, we have verified that it would be  immediate to obtain positive-definite 
(and monotonically decreasing) 
solutions if we could reverse the overall sign of the equation (or, equivalently, the sign of the kernel).
Thus, the existence of a  negative region in the kernel is a necessary condition for obtaining physical solutions; 
a positive-definite kernel, would exclude immediately such a possibility.
Of course, as we will see shortly, this condition is far from sufficient. 

\begin{figure}[!t]
\hspace{-1cm}\includegraphics[scale=1]{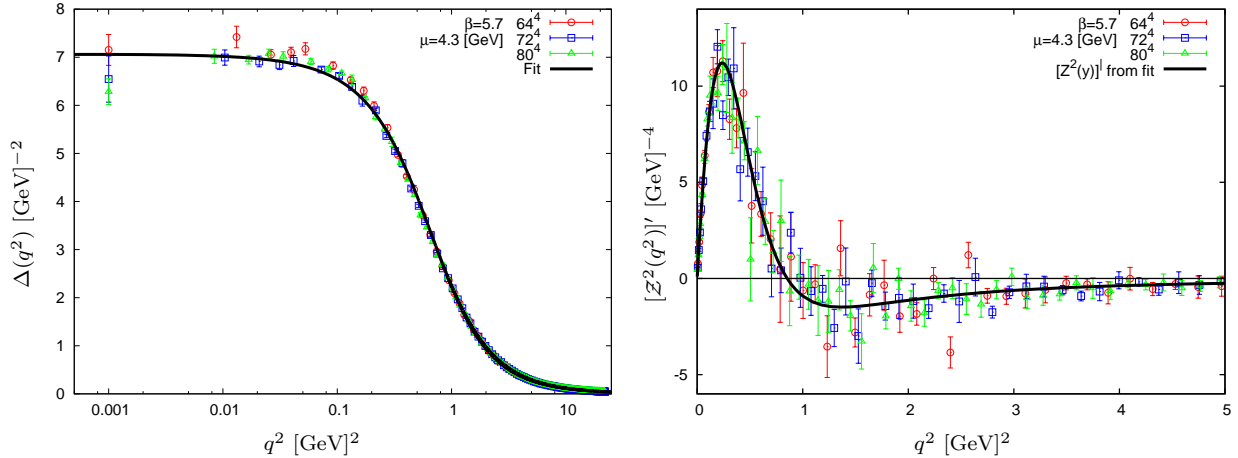}
\caption{(color online). ({\it Left panel}.) Lattice data for the (quenched) $SU(3)$ gluon propagator renormalized  at $\mu=4.3$ GeV, shown together with the log fit 
of~\cite{Aguilar:2010cn,Aguilar:2011ux,Aguilar:2010gm}. ({\it Right panel.}) The one-loop dressed kernel derived from the lattice data, compared to the one obtained from the propagator fit.}
\label{gprop}
\end{figure}

Specifically, using 
the lattice data for $\Delta(q^2)$ corresponding to a $SU(3)$ quenched lattice simulation~\cite{Bogolubsky:2007ud},  
one can explicitly verify whether or not (and exactly how) 
the necessary condition described above is satisfied. These lattice data are plotted in \fig{gprop} (left panel); on  the same figure we also  
plot a fit, whose explicit 
functional form may be
found in various recent articles~\cite{Aguilar:2010cn,Aguilar:2011ux,Aguilar:2010gm}).  
On the right panel of Fig.~\ref{gprop} we then show the corresponding one-loop dressed kernel ${\cal K}_1$,  
calculated directly from the lattice data\footnote{The kernel in this case is obtained by first calculating 
from the raw lattice data $(y_i,\Delta(y_i))$ the dressing function squared data $(y_i,{\cal Z}^2(y_i))$; 
next the derivative data are calculated as a simple first order central difference, \ie the data plotted are $(\frac{y_{i+1}+y_i}2,\frac{{\cal Z}^2(y_{i+1})-{\cal Z}^2(y_{i})}{y_{i+1}-y_i})$. Finally, errors are calculated through error propagation.} 
and then also using the aforementioned fit. 
One clearly observes the zero crossing of the kernel (at $q^2\sim0.8$ GeV$^2$) and the corresponding negative region. 

The propagator fit shown in~\fig{gprop}  can then be used to construct
the   full   kernel  appearing   in~\1eq{me-gen},   when  $C=0$;   the
corresponding full  integral equation can  be then studied  by means of 
the algorithm explained above (for  the ghost dressing function $F$, we
use  a continuous  interpolator of  the corresponding  $SU(3)$ lattice
data).  It turns  out, however,  that  the solutions  obtained have 
oscillatory  behavior, and display  large negative  regions, ultimately
voiding them of any physical meaning. 

Evidently, the negative region furnished by the kernel is not sufficiently deep, or it is not located
in the optimum momentum region, to counteract the effect of the overall minus sign. Clearly, if physical solutions 
are to be found, the full functional form of the kernel must be modified. As we will see in the next subsection, this type of 
appropriate modification is indeed implemented dynamically, when the two-loop corrections are included.

Finally, in order to avoid any confusion related to the conclusions of this subsection and the findings of~\cite{Aguilar:2011ux}, 
let us remind the reader  
that the equation solved in~\cite{Aguilar:2011ux} does not coincide with the one solved here; therefore, the 
(non-monotonic) solutions found in~\cite{Aguilar:2011ux}, do not correspond to solutions of the present integral equation. 

\begin{figure}[!t]
\mbox{}\hspace{-1.2cm}\includegraphics[scale=.9]{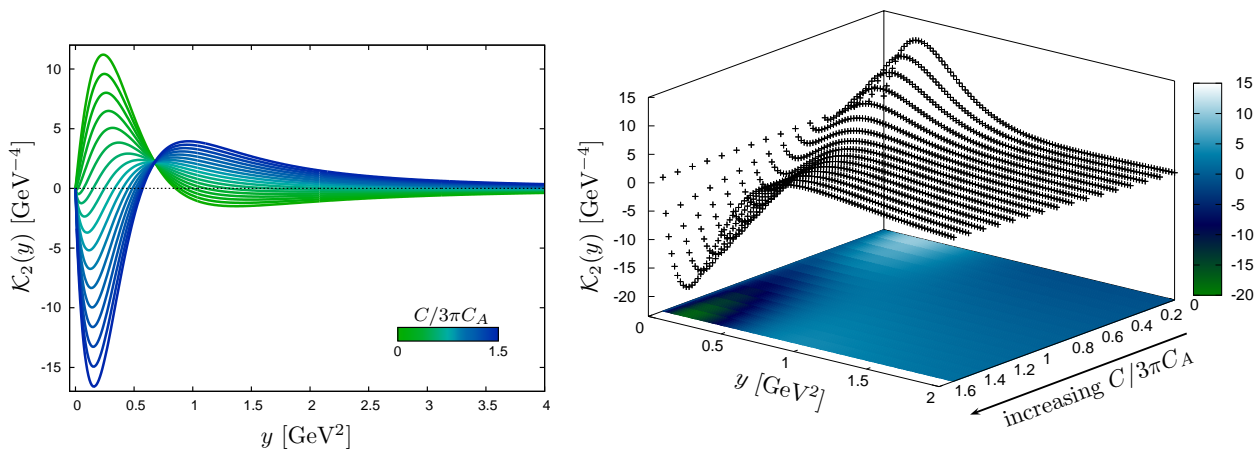}   
\caption{\label{kernel2d-3d}(color online). Modification of the shape of the two-loop dressed kernel ${\cal K}_2(y)$ with varying $C$.  
As the latter parameter increases the kernel effectively reverses its sign showing a deep negative well in the low momenta region.}  
\end{figure}

\subsection{The two-loop dressed case: finding physical solutions}

Let us now turn on again the two-loop dressed contributions, by setting $C>0$. Considering again the $x\to0$ limit first, 
one has in this case
\be
m^2(0)=-\frac{3\pi}{4}\lambda\, F(0)\int_0^\infty\! \diff y m^2(y){\cal K}_{\rm 2}(y);\qquad  {\cal K}_{\rm 2}(y)=\left\{\left[1-2CY(y)\right]{\cal Z}^2(y)\right\}'.
\label{fullmass0}
\ee

Since, as already mentioned, $C$ will be treated as an independent parameter, one can study how the shape of the 
kernel ${\cal K}_2$ changes as $C$ is varied. As can be seen in~ \fig{kernel2d-3d}, when $C=0$ one is back to 
the one-loop dressed kernel ${\cal K}_1$ of the previous subsection. As $C$ increases ${\cal K}_2$ displays 
a less pronounced positive (respectively negative) peak in the small (respectively large) momenta region. 
Next, for $C\gtrsim0.37$, a small negative region starts to appear in the IR, which rapidly becomes 
a deep negative well for $y\lesssim0.6$, with ${\cal K}_2$ becoming positive for higher momenta\footnote{It is also interesting to notice 
that all the curves meet at a common point $y_*$, which is determined by the general condition
$$
\left.\frac{\diff}{\diff y}\left[\log\left({\cal Z}^2(y)\log\frac{y}{\mu^2}\right)\right]\right|_{y=y_*}=0, 
$$  
with $y_*\simeq0.67$ GeV$^2$.}.
Therefore, we see that the addition of the two-loop dressed contributions counteracts the effect of 
the overall minus sign of the integral equation, by effectively achieving 
a sign reversal of the kernel (roughly speaking, one has ${\cal K}_2\approx-{\cal K}_1$).
When this  analysis is combined  with the knowledge gathered  from the
one-loop  dressed case,  one concludes  that there  exists  a critical
value  $\overline{C}$, such  that,  if $C>\overline{C}$  \1eq{fullmass0}
will display  at least one physical  monotonically decreasing solution
for a suitable value of  the strong coupling $\alpha_s$.

\begin{figure}
\mbox{}\hspace{1.1cm}\includegraphics[scale=0.725]{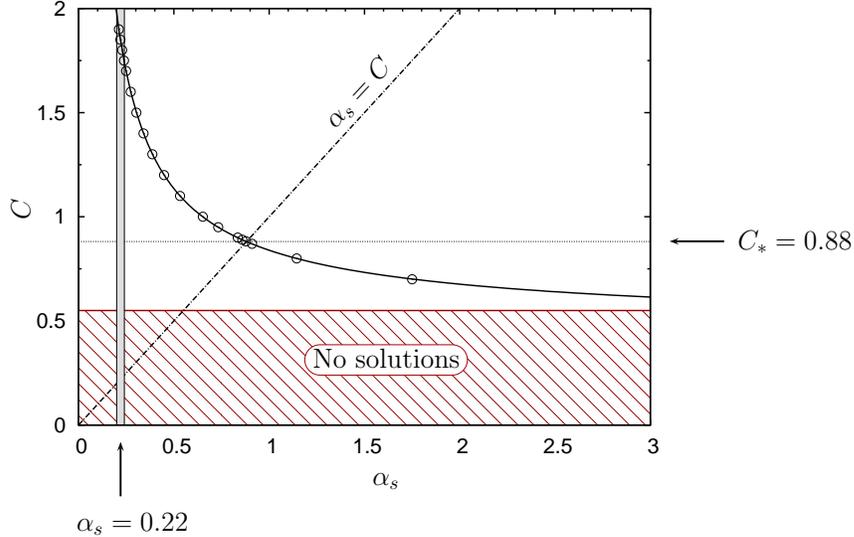}  
\caption{\label{Cvsalpha}(color online). The curve described by the set of the pairs $(C,\alpha_s)$ for which one finds physical solutions to the full mass equation~\noeq{me-gen}. The curve starts from the critical value $\overline{C}\approx0.56$ above which exactly one monotonically decreasing solution exists (below $\overline{C}$ there are no solutions). The value $C_*=\alpha_s(\approx0.88)$ corresponds to the case in which $Y$ is kept at its lowest order perturbative value. Finally, the gray vertical band represents the value for the quenched strong coupling obtained from the four-loop MOM calculation of~\cite{Boucaud:2005rm} renormalized at $\mu=4.3$ GeV ($\alpha_s=0.22$) with a customary $10\%$ error; the matching values of $C$ are between $1.8$ and $1.9$.}
\end{figure} 
 
In  order to  see  if the  picture  sketched above  is confirmed  when $x\neq0$,  one can  study numerically  the solutions  of the  full mass equation~\noeq{me-gen}  following  the   algorithm  described  at  the beginning of the section (\fig{Cvsalpha}). 
Specifically, as shown in~\fig{Cvsalpha}, there is a 
continuous curve formed by the pairs ($C,\alpha_s$), for which one finds physical solutions. Indeed, for small values of $C$ one
has that all  eigenvalues  $\lambda$  are  either negative  or
complex, and  no solution exists;  this absence of  solutions persists until the  critical value $\overline{C}\approx0.56$  is reached, after which  one   finds  exactly  one   monotonically  decreasing  solution corresponding  to the smallest  positive eigenvalue  $\lambda_s$ (with bigger  positive eigenvalues giving  rise to  oscillating non-physical solutions).  However,  for values  up  to  $C\approx0.8$ the  coupling needed  to get  the corresponding  running mass  is of  ${\cal O}(1)$, while for the quenched case  the expected coupling from the 4-loop (momentum subtraction) calculation   is  $\alpha_s=0.22$   at  $\mu=4.3$   GeV~\cite{Boucaud:2005rm}.  This latter value is obtained for $C\approx1.8$ -- $1.9$, whereas for $C\approx0.88$ one finds the solution to~\1eq{me-gen} for the 
lowest-order perturbative value of the coefficient~\noeq{theC}. In general one observes, as expected, that as $C$ is increased, $\alpha_s$ decreases, \eg for  $C=1.1$, $1.3$, $1.5$, and $1.7$  one  obtains solutions  corresponding  to  the strong coupling values $\alpha_s\approx0.53$, $0.39$, $0.30$, and $0.25$, respectively.  

\begin{figure}
\mbox{}\hspace{-1cm}\includegraphics[scale=0.725]{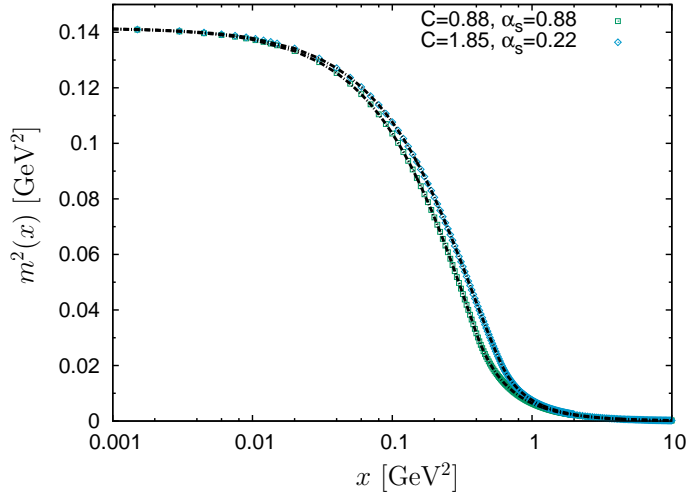} 
\caption{\label{full-sols}(color online). Typical monotonically decreasing solution of the mass equation~\noeq{me-gen}. 
The case shown have been obtained for the special values $C=0.88$ and $1.85$, corresponding value of the coupling $\alpha_s\approx0.88$ and $0.22$, respectively. The solutions have been normalized so that at zero they match the corresponding (Landau gauge) lattice value $\Delta^{-1}(0)\approx0.141$ GeV$^2$.}
\end{figure}

In~\fig{full-sols}  we plot the solutions for the most representative $C$ values, \ie $C=0.88$ and $C=1.85$ (corresponding to, as already said,  $\alpha_s\approx0.88$ and $0.22$, respectively); notice that we have used the linearity of the equation to normalize the solution in such a way that the mass at zero coincides with the IR saturating value found in lattice (Landau gauge) quenched simulations~\cite{Bogolubsky:2009dc}, or $m^2(0)=\Delta^{-1}(0)\approx0.141$ GeV$^2$. 
As can be readily appreciated, the masses obtained 
display the basic qualitative features expected on  
general field-theoretic considerations and employed in numerous phenomenological  studies; 
in particular, they are monotonically decreasing functions of the momentum, 
and vanish rather rapidly in the ultraviolet~\cite{Cornwall:1981zr,Lavelle:1991ve,Aguilar:2007ie}. 
It would seem, therefore, that the all-order analysis presented here, puts 
the entire concept of the gluon mass, and a variety of fundamental properties ascribed to it, on a  
solid first-principle basis.  

\section{\label{concl}Conclusions} 

In the  present work we  have derived the {\it  full} non-perturbative
integral equation that governs the momentum evolution of the dynamical
gluon  mass.   This  has  been  accomplished  by  determining  through
detailed calculations the mass-related contributions of the ``two-loop
dressed''  graphs that  appear in  the  SDE of  the gluon  propagator,
within the  PT-BFM framework.  Thus, the complete  equation emerges by
including  these  latter  contributions  to those  obtained  from  the
one-loop dressed analysis presented in~\cite{Aguilar:2011ux}.

We have explained in detail the methodology that allows for a systematic 
and expeditious identification of the parts of the SDE that contribute to the 
mass equation. In particular, given the manifest transversality of the full gluon self-energy, 
we have focused on the $q_{\mu}q_{\nu}$ component of the answer, 
thus avoiding a number of technical subtleties related to the use of the ``seagull identity''. 
Then, the relevant contributions stem entirely from the 
parts of the diagrams that involve the 
special nonperturbative vertices associated with the Schwinger mechanism
(generically denoted by $V$), which  
are longitudinally coupled, contain massless poles, and satisfy relatively simple WIs. 
In fact, the fully longitudinal nature of these vertices, coupled to the 
use of the Landau gauge, obviates the need of their explicit construction;
indeed, the desired result may be obtained simply by resorting to the WIs that these  
vertices obey, without the need to invoke their explicit closed form. 

It turns out that the  inclusion of two-loop dressed contributions has
a profound  impact  on  the  nature of the mass equation, already at 
the qualitative level.   
In fact, the kernel  of the integral equation acquires
an  extra  term, with respect to the one-loop case~\cite{Aguilar:2011ux}, 
whose presence modifies the nature of the obtained solutions.
Specifically, even within the simplest one-loop approximation, 
this additional term allows for the appearance of 
positive-definite and monotonically decreasing solutions for the gluon mass $m^2(q^2)$. 
A detailed numerical investigation of the resulting homogeneous linear integral equation 
was performed, and the dependence of the solutions on the value of the 
strong coupling $\alpha_s$ and an appropriately introduced parameter $C$ expressing our uncertainty on the two-loop term $Y(k^2)$, has been studied. 

Our analysis reveals that there is a critical $C$ value ($\overline{C}\approx0.56)$  below which no physically admissible  solutions may be found. On the other hand, perfectly acceptable solutions are obtained above this value, with the solution corresponding to the strong coupling $\alpha_s=0.22$, predicted at $\mu=4.3$ GeV in the MOM scheme, obtained for $C=1.85$. Within the lowest-order approximation employed for calculating the two-loop dressed term $Y$, a monotonically decreasing mass is obtained for the strong coupling value $\alpha_s\approx0.88$. The fact that already, within this simple approximation for $Y$, one is able to obtain a solution of the full mass equation for a relatively modest size strong coupling should be regarded as a success of the entire PT-BFM framework. 
 
Given the importance of the term  $\Y(k^2)$ for this entire construction, it would  certainly be important to determine its structure beyond the one-loop approximation used. 
As already mentioned in Section~\ref{fullmasseq}, 
this can be accomplished, to a certain extent, by making use of the Ball and Chiu Ansatz 
for the three-gluon vertex~\cite{Ball:1980ax}. 
An alternative might be lattice itself, given that 
the three-gluon vertex is \n{i} computed in the Landau gauge, and \n{ii} it is 
naturally contracted by three transverse projection operators [see \2eqs{defY}{fnal}]. Interestingly enough, the resulting structure $PPP\Gamma$ is precisely the one that is accessible to Landau gauge lattice simulations of the gluon three-point function~\cite{Cucchieri:2008qm}. 

The full mass equation, even with the approximate version of $\Y(k^2)$, 
provides a natural starting point for calculating reliably 
the effect that the inclusion of light quark flavors might have on 
the form of the gluon propagator in general, and on the gluon mass in particular.   
Specifically, recent studies based on SDEs~\cite{Aguilar:2012rz} as well as lattice simulations for $N_f=2$, and $N_f=2+1+1$~\cite{Ayala:2012pb}, appear to 
converge to the conclusion that the quark effects tend to suppress the gluon propagator in the intermediate and low-momentum region, causing it to saturate at a slightly lower point,  compared to the saturation point
observed in the quenched case. Should this picture persists further scrutiny, it would seem to indicate, among other things, that the gluon mass increases 
in the presence of light quarks. It would be, therefore, important to obtain a concrete estimate for the behavior of the gluon mass in the presence of dynamical quarks and especially its value at the origin, \ie  $m^2(0)$. 
We hope to be able to report progress in this direction in the near future.

\begin{acknowledgments}     

Useful discussions with A. C. Aguilar are gratefully acknowledged. The research of D. I. and J. P.  is supported by the Spanish MEYC under Grant No. FPA2011-23596.   
 
\end{acknowledgments}


\end{document}